\documentclass[journal=tches]{iacrtrans}

\usepackage{lipsum} 
\usepackage{graphicx}
\usepackage{colortbl}
\usepackage[colorinlistoftodos]{todonotes}
\usepackage{amsmath,mathtools}
\usepackage{hyperref}
\usepackage{graphicx}
\usepackage{multirow}
\usepackage{pifont}
\newcommand\numberthis{\addtocounter{equation}{1}\tag{\theequation}}

\usepackage{threeparttable}
\usepackage[ruled,vlined,linesnumbered]{algorithm2e}
\usepackage{setspace}
\usepackage[skip=5pt]{caption}
\usepackage{subfig}
\usepackage{enumitem}
\usepackage{algorithmicx}
\usepackage{color}
\definecolor{mygray}{gray}{0.85}
\newcommand\newversion[1]{{\color{black} {#1}}}
\newcommand\majorrevision[1]{{\color{black}{#1}}}
\definecolor{mygreen}{RGB}{0,100,0}
\newcommand\ndss[1]{{\color{black} {#1}}}
\newcommand\minr[1]{{\color{black} {#1}}}

\usepackage{fixltx2e}
\def\SPSB#1#2{\rlap{\textsuperscript{\textcolor{black}{#1}}}\SB{#2}}
\def\SP#1{\textsuperscript{\textcolor{black}{#1}}}
\def\SB#1{\textsubscript{\textcolor{black}{#1}}}

\DeclareMathOperator*{\argmax}{arg\,max}
\newtheorem{Def}{Definition}

\graphicspath{{Figures/}}
\DeclareGraphicsExtensions{.pdf,.png,.jpg,.eps} 

\algnewcommand{\IfThenElse}[3]{
  \State \algorithmicif\ #1\ \algorithmicthen\ #2\ \algorithmicelse\ #3}

\author{Rabin Y. Acharya\inst{1} \and Fatemeh Ganji\inst{2} \and Domenic Forte\inst{1}}
\institute{
  University of Florida, Gainesville, USA, \email{rabin.acharya@ufl.edu, dforte@ece.ufl.edu}
  \and
  Worcester Polytechnic Institute, Worcester, USA, \email{fganji@wpi.edu}
}

\title{Information Theory-based Evolution of Neural Networks for Side-channel Analysis}

\begin{document}

\maketitle


\begin{abstract}
Profiled side-channel analysis (SCA) leverages leakage from cryptographic implementations to extract the secret key. 
When combined with advanced methods in neural networks (NNs), profiled SCA can successfully attack even those crypto-cores assumed to be protected against SCA. 
Despite the rise in the number of studies devoted to NN-based SCA, a range of questions has remained unanswered, namely: how to choose an NN with an adequate configuration, how to tune the NN's hyperparameters, when to stop the training, etc. Our proposed approach, ``InfoNEAT,'' tackles these issues in a natural way. 
InfoNEAT relies on the concept of neural structure search, enhanced by information-theoretic metrics to guide the evolution, halt it with novel stopping criteria, and improve time-complexity and memory footprint. 
The performance of InfoNEAT is evaluated by applying it to publicly available datasets composed of real side-channel measurements. 
In addition to the considerable advantages regarding the automated configuration of NNs, InfoNEAT demonstrates significant improvements over other approaches for \ndss{effective key recovery in terms of the number of epochs (e.g.,$\times$6 faster) and the number of attack traces compared to both MLPs and CNNs (e.g., up to 1000s fewer traces to break a device) as well as a reduction in the number of trainable parameters compared to MLPs (e.g., by the factor of up to 32). 
Furthermore, through experiments, it is demonstrated that InfoNEAT's models are robust against noise and desynchronization in traces. }
\end{abstract}

\section{Introduction}\label{sec:introduction}
\sloppypar{
While promising candidates have been proposed in the literature to deal with secure generation and key storage, the issue with secure execution continues to exist. 
Secure execution is a challenging goal to attain due to the leakage of information from implementations, often referred to as a side-channel. 
Attacks leveraging such secret-key leakages are strong in the sense that they have broken the security of various real-world devices. 
The involvement of standardization and certification bodies in the
activities related to the development of tools for leakage detection and side-channel analysis (SCA) further highlights the importance of this matter, see, e.g.,~\cite{NIST2020,EU2020}.  

Prominent examples of side-channels include execution time, power consumption, and electromagnetic (EM) radiations collected from a target, e.g., a device embodying a cryptographic module such as an advanced encryption standard (AES) crypto-core. 
To analyze such side-channels, profiled attacks are common practice due to their effectiveness~\cite{ascad_paper}. 
In the first phase of this attack (so-called profiling), the leakage from an open copy of the target device, is modeled under a controlled condition (e.g., known secret keys) cf.~\cite{chari2002template,hospodar2011machine,lerman2015machine,martinasek2013optimization}. 
Traditionally, such a model has been obtained by characterizing the leakages precisely through statistical techniques, e.g., linear regression~\cite{schindler2005stochastic,doget2011univariate}. 
Afterward, in the second phase (so-called attack), this model is used to launch a key-recovery attack on the target device. 
These phases are in line with the specifics of machine learning (ML) tasks in the sense that the profiling and attack steps correspond respectively to the training and testing sub-tasks in the context of supervised ML. 
Interestingly enough, ML-enhanced SCA can defeat not only unprotected, see e.g.,~\cite{lerman2015template,heuser2012intelligent,hospodar2011machine}, but also protected cryptographic implementations~\cite{lerman2015machine,gilmore2015neural,cagli2017convolutional,related_works:SCA_metrics,kim2019make,related_works:automated_hyperparameter_tuning}. 
Furthermore, when only a limited number of traces with noise is available, ML-enhanced SCA approaches outperform the template attack~\cite{chari2002template}, known to be optimal from an information-theoretic perspective if a large enough number of traces are available~\cite{related_works:SCA_metrics,bhasin2020mind}. 
The scenario, where only a few noisy traces are available, is of great importance as it reflects a more realistic scenario.  

In this regard, it is not surprising that deep learning and NNs are now playing an active role in the SCA literature~\cite{hettwer2020applications,picek2021sok}. 
In particular, it has been demonstrated that NNs can further defeat some countermeasures designed to protect a cryptographic implementation. Specifically, the jitter-based misalignments in the side-channel traces, i.e., creating an array of asynchronous measurements, cannot stop an attacker from launching SCA through NNs~\cite{cagli2017convolutional,ascad_paper}. 
Even masked implementations, with countermeasures that randomize the intermediate values, can be successfully attacked by NNs~\cite{kim2019make,related_works:automated_hyperparameter_tuning,related_works:SCA_metrics}. 
\ndss{
In fact, methodologies, techniques, and solutions have been proposed to help evaluators at the analysis stage of NN-enabled SCA cf.~\cite{rioja2021uncertainty}. 
An expensive profiling step can account for the cost of NN-enabled SCA~\cite{azouaoui2020systematic}, where the NN configuration and hyperparameters should be determined. 
Therefore, although it seems tempting to consider NN-enabled SCA a step toward automating SCA,} the design of NNs needs to become less expert-dependant. 
More concretely, these questions need to be answered, as they have remained partially open so far: 
\emph{Which configurations and hyperparameter combinations should be used?  
When can the training process be stopped to achieve better generalization in the attack phase?}



\vspace{1ex}

\noindent\textbf{InfoNEAT: } 
\ndss{Our paper introduces InfoNEAT that is the \emph{first} NN-enhanced SCA taking advantage of neural architecture search (NAS) at its \textbf{\emph{full}} capacity (see Section~\ref{sec:related_work_hyper} for comparison with the most relevant approaches).}
InfoNEAT is a novel algorithm that revolves around the notion of the evolution of NNs, so-called ``neuroevolution.''
The cornerstone of this concept is to evolve the \textit{configuration of multiple networks (so-called augmenting topologies), and simultaneously tune their hyperparameters and parameters}.  
Specifically, it has been demonstrated that although NN-based attacks can extract the secret from even protected implementations, hyperparameter tuning imposes serious challenges to the application of NNs for SCA~\cite{related_works:automated_hyperparameter_tuning,kim2019make,zaid2020methodology}. 
Interestingly, although neuroevolution can be employed in various domains, InfoNEAT is tailored to the specific requirements of SCA. 
The workflow of InfoNEAT\footnote{The source code is available in the Github https://github.com/rachary00/InfoNEAT} is similar to common profiled attacks, see Figure~\ref{fig:infoneat_overview}. 
For existing NN-based attacks, in the training process, selecting the adequate configuration of the network and tuning the hyperparameters are often based on trial and error (see Section~\ref{sec:related_works} for more details). 
InfoNEAT, on the contrary, evolves various NNs and their hyperparameters and eventually delivers the one selected based on sound information-theoretic metrics. 
Regarding our \textbf{contributions} listed below and expounded upon in the paper, InfoNEAT is particularly powerful for SCA. 

\textbf{(1) InfoNEAT tailors NAS for SCA.} 
\ndss{InfoNEAT takes advantage of the entire range of modern NN design elements, including hyperparameter tuning and multi-branch configuration. 
In doing so, the irregular topology of NNs (compared to the structure of multi-layer perceptrons--MLPs) automatically discovered by NEAT is one of the most important aspects of our framework, which can lead to the robustness against desynchronization as defined in SCA (see Section~\ref{sec:results_sca_metrics}).}
The original NEAT algorithm~\cite{stanley2002evolving} cannot cope with complex multi-class classification tasks, specifically when many output neurons are involved.
To address this, InfoNEAT employs One-vs-All (sometimes called ``One-vs-Rest'') multi-class classification followed by stacking ensemble learning. 
\newversion{This is indeed an elegant and effective method to \emph{combine} classifiers for all sub-keys, although it has not been  studied in the SCA-related literature. }
This is one of the novel aspects of InfoNEAT that conforms with the specification of SCA\ndss{, where the number of classes is much larger than what has been considered in neuroevolution-related literature (see Section~\ref{sec:methodology_training} for more details). }


\textbf{(2) Criteria for stopping the training/evolution processes in InfoNEAT.} 
\ndss{When introducing NEAT, defining such criteria has been identified as a challenge~\cite{stanley2002evolving}, however this issue has not been resolved, but rather handled through trial and error.}
It is evident that evolving NNs could be a time-consuming and memory-hungry process as done by the original NEAT algorithm.
To address this, InfoNEAT is equipped with mechanisms relying on information theory.
Intuitively, the information-theoretic conditions indicate whether adding a new connection/node can help improve InfoNEAT's output.
If not, the evolution process is stopped. 
Moreover, the proposed criteria are useful to guide the evolution process by selecting the best NN among ones evolved by InfoNEAT.
\ndss{For SCA, the implication of this is that stopping early enough avoids overfitting and obtains better generalization to unseen testing traces.}

\textbf{(3)  Key-recovery attacks using InfoNEAT.} 
Last but not least, we evaluate the effectiveness of InfoNEAT against state-of-the-art (SOTA) side-channel trace databases and ML models (MLPs and CNNs) that have become the standard benchmarking system for SCA. 
InfoNEAT achieves competitive results compared to its counterparts, see Section~\ref{sec:results_sca_metrics}, without relying on additional techniques (e.g., validation step) employed against challenging datasets in SOTA studies. 
InfoNEAT can also enhance the memory and time efficiency of SCA by reducing the number of trainable parameters in NNs and  epochs\footnote{The number of epochs is the number of times a learning algorithm sees the entire training dataset.}, \ndss{e.g., by factors of up to 32 and 6, respectively, when compared to the approach in~\cite{related_works:automated_hyperparameter_tuning} that considers hyperparameter tuning for MLPs, counterparts of NNs delivered by InfoNEAT (for comparison with other SOTA attacks, see Section~\ref{sec:results}).}


\begin{figure}[t]
\centering
\includegraphics[width=0.6\columnwidth]{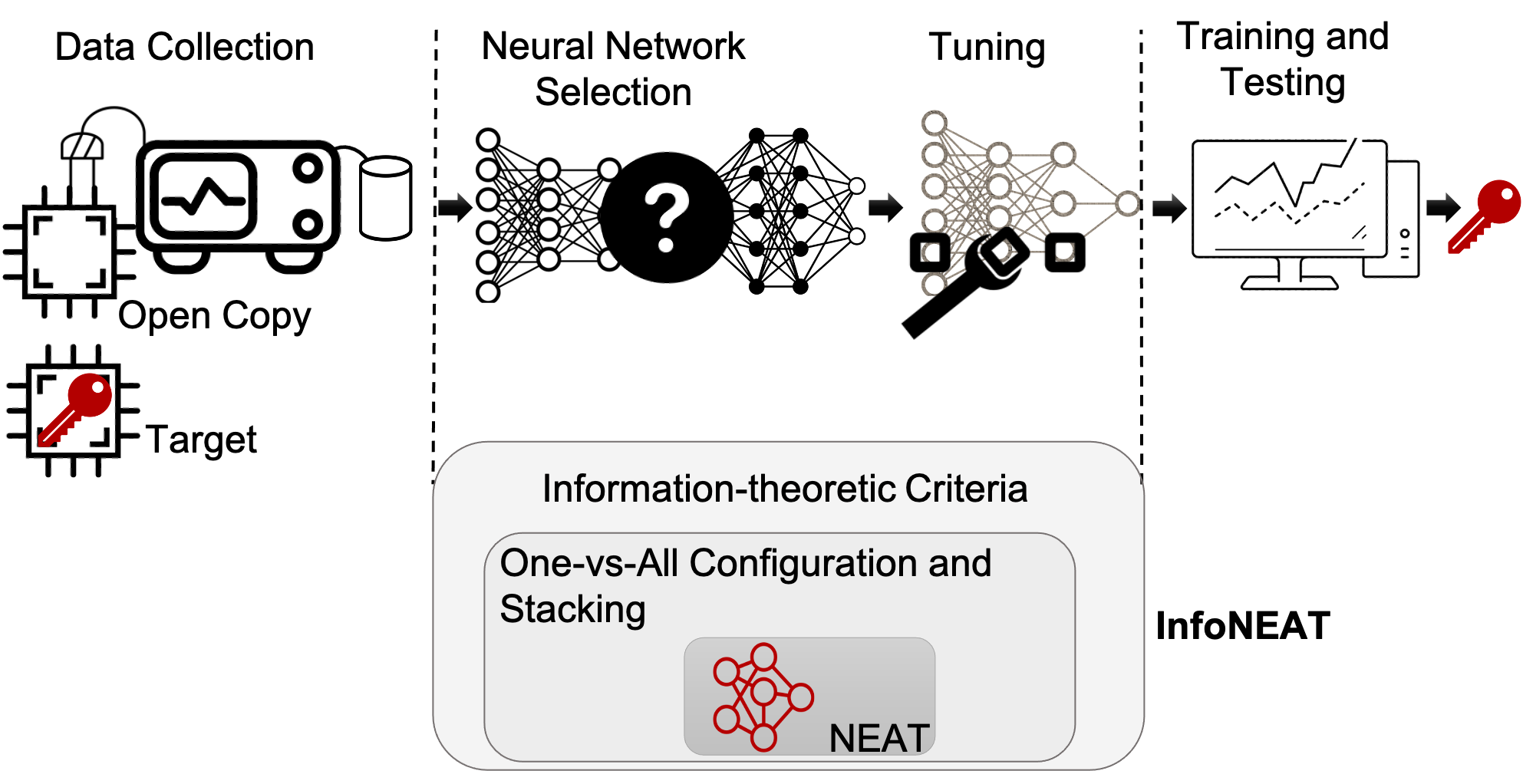} 
\caption{Overview of the proposed ``InfoNEAT'' which extends the NEAT algorithm to meet SCA requirements.
Compared to conventional ML-enabled SCA, network selection and hyperparameter tuning are handled automatically.}
\label{fig:infoneat_overview}
\end{figure}

\noindent\textbf{Outline:}
The rest of the paper is organized as follows.
Section~\ref{sec:related_works} gives a brief overview of the most relevant studies in the literature.
In Section~\ref{sec:background}, profiled SCA along with our notations and mathematical principles are discussed.
Afterward, in Section~\ref{sec:neat}, we introduce the NEAT algorithm as a building block of our proposed InfoNEAT algorithm described in Section~\ref{sec:infoneat}.
This is then followed by Section~\ref{sec:results}, where the results of our experiments are presented and discussed.
Finally, Section~\ref{sec:conclusion} concludes the paper. 
}

\section{Related Work}\label{sec:related_works}
Machine learning techniques have become quite popular and successful in profiled SCA in recent years.
Among them, multilayer perceptron (MLP) and convolutional neural networks (CNNs) are the two most widely used~\cite{related_works:MLP_weissbart,related_works:CNN_performance,related_works:one_trace_CNN,hettwer2020applications}. 
\ndss{Leaving aside the selection of network type, tuning the hyperparameters of NNs remains an issue of concern as inappropriate choices may} lead to \textit{overfitting}.  
In this section, we explain these issues as well as the relevant works in more detail. A summary of the related works, as well as the comparison with InfoNEAT, are also included in Table~\ref{tab:related_works} (see Appendix~A).

\subsection{Type of Neural Networks} 
\ndss{Similar to other ML tasks, selecting the model, e.g., NNs with proper configuration, is the first step and maybe, one of the most challenging ones.  
When combining CNNs with data augmentation techniques, one could launch successful SCA against even protected designs~\cite{cagli2017convolutional,pu2017trace}. 
As an example, a recent study has applied multi-scale CNN along with various data transformations (e.g., phase-only correlation, principal component analysis--PCA, alignment methods) to allow a better attack performance~\cite{won2021back}. 
In this regard, it has been demonstrated that advantages offered by CNNs can be most enjoyed when the level of noise is small and the number of measurements and features is high~\cite{related_works:CNN_performance}. 
From another perspective, one of the reasons behind the success of CNNs in image classification is maintaining the ordering of features. 
Interestingly enough, this feature could be less important for SCA, cf.~\cite{related_works:CNN_performance}. } 
In line with this, in~\cite{hettwer2020applications,related_works:plaintext}, the authors have mentioned that CNNs could outperform other NN-based attack techniques because of their effectiveness with raw data especially in the presence of jitter or desynchronization.
This comparison, however, should be considered carefully in the field of SCA as some of the techniques have traditionally only considered ML metrics to evaluate their models~\cite{related_works:SCA_metrics,related_works:CNN_performance}. 
This, of course, does not rule out the adoption of CNNs in this domain but opens up new avenues for other NNs. 
From this perspective, the authors in~\cite{related_works:SCA_metrics} have shown that when considering SCA-related metrics, both models (CNN and MLP) perform similarly well, especially when combined with SMOTE (Synthetic Minority Oversampling Technique). 
Picek et al. in~\cite{related_works:CNN_performance} showed that MLPs equipped with XGBoost outperform CNN when considering pure SCA metrics such as the guessing entropy. 

\ndss{Another feature of CNNs that could be beneficial to SCA is their implicit feature selection part. 
This has been well-studied in~\cite{Wouters_Arribas_Gierlichs_Preneel_2020}, where the first convolutional layer of the SOTA CNN configured by Zaid et al.~\cite{zaid2020methodology} is replaced by a classical preprocessing technique to reduce the model's complexity effectively. 
Such combinations of preprocessing and CNNs can reduce the cost of training CNNs. 
In some scenarios, this cost could be justified if the performance of the attack is improved, e.g., when a trained CNN (e.g., VGG16) is used for SCA~\cite{ascad_paper,masure2021side,related_works:plaintext}. 
In doing so, an interesting case is discussed in~\cite{related_works:plaintext}, where a model trained on a dataset is applied against the desynchronized traces in that dataset. 
To our knowledge, MLPs have not been considered in this case; therefore, a careful cost assessment of training MLPs and CNNs under such circumstances is needed (see Section~\ref{sec:results_sca} for our results).   }
This discussion leads us to one of the major issues in the SCA community, \textit{generalization}.

\subsection{The Issue of Generalization}
Generalization is the ability of a trained SCA model to adapt well to new, unseen data. 
\majorrevision{In our framework, in line with other relevant studies, generalization from training to test sub-datasets is considered.} 
In this context, the authors in~\cite{related_works:explainability_SCA} demonstrate attacks on different SCA datasets and show how MLPs are internally the same despite changing the device or the key because MLPs are considered universal approximators. 
Furthermore, even small MLP networks were shown to roughly learn the same function without overfitting and generalize to different datasets~\cite{related_works:least_num_traces,related_works:explainability_SCA}. Generalization has also been addressed in~\cite{related_works:ensemble_SCA}, where the authors show how ensembles of models based on averaged class probabilities can further improve the network generalization for various datasets. 
For this purpose,~\cite{related_works:ensemble_SCA} has applied the bagging ensemble technique to combine models trained on all classes. 
In contrast to that, the model selection performed in the context of InfoNEAT is similar to the ``bucket of models'' ensemble method enhanced with stacking to combine sub-models generated per class (see Section~\ref{sec:methodology_training} for more details). 
Accordingly, InfoNEAT is further boosted to improve the generalization among various classes. 
Tuning of the \textit{hyperparameters} also has a direct impact on generalization. 
Next, we consider the most relevant literature devoted to this matter. 

\ndss{\subsection{Hyperparameter Tuning and NAS}\label{sec:related_work_hyper}
Before reviewing the studies on designing NNs for SCA, it is useful to distinguish between hyperparameter optimization and NAS. 
A NAS approach attempts to discover neural architectures by searching a specific space. 
The following items typically parameterize a search space:  
(a) The (maximum) number of layers that can be possibly unbounded; 
(b) the type of operation for each layer, e.g., pooling and convolution; 
(c) hyperparameters associated with the operation per layer, e.g., the number of filters, kernel size and strides for a convolutional layer, the number of neurons in a layer of MLPs, and the number of epochs; 
(d) finally, more advanced methods incorporate multiple branches and skip connections~\cite{elsken2019neural}. 
InfoNEAT is the \emph{first} NN-enhanced SCA that considers all these items through neuroevolution. Nevertheless, the most relevant studies are discussed below.


In the same vein as other ML tasks, for SCA, simpler models could be preferred when available
resources are limited; however, the attack performance should not be compromised. 
The trade-off between these has been studied in the SCA-related literature. 
Conventionally, it has been thought that when attacks are harder to mount, e.g., in cases with noisy traces and countermeasure-protected designs, small (not using too many layers) NNs could not achieve the performance desired for a successful SCA; nonetheless, it has been shown that having a deeper network is not necessarily an advantage~\cite{van2020learning,related_works:automated_hyperparameter_tuning}. 
Hence, for SCA, one of the hyperparameters to tune is the maximum number of layers. }
Besides the number of layers (a), the type of operation for each layer and their hyperparameters (b and c) have been considered in recent works. 
Random search within pre-defined ranges~\cite{related_works:ensemble_SCA,related_works:automated_hyperparameter_tuning}, and grid search\cite{related_works:MLP_weissbart} have been considered to tune the hyperparameters.  
An example of hyperparameter tuning for choosing the type of operations is given in~\cite{perin2020influence,kerkhof2021no}, where numerous choices for optimizers and loss functions have been explored. 
In~\cite{related_works:MI_SCA}, the authors suggest tuning hyperparameters, while trying to prevent underfitting or overfitting by defining stopping criteria (rather than a pre-defined number of epochs) by relying on the mutual information between output layer activation functions (Softmax) and the data labels (for a comparison with InfoNEAT, see Section~\ref{sec:method_info}).  

\majorrevision{
Arguably, for SCA, Zaid et al.~\cite{zaid2020methodology} proposed the first methodology devoted to the explainability and interpretability of model hyperparameters.  
For this, visualization techniques have been applied to determine the impact of CNN model hyperparameters related to the convolutional part, e.g., the number of filters. 
With regard to their impact, one attempts to find a suitable architecture with a minimized complexity, cf.~\cite{zaid2020methodology}. }
\ndss{Nevertheless, their method suffers from issues discussed in~\cite{Wouters_Arribas_Gierlichs_Preneel_2020} where the authors improve the configuration of CNNs and significantly reduce their size without compromising the attack performance. 
Similarly, towards making the NN-enhanced SCA more automated,}~\cite{related_works:hyperparameter_tuning_RL} uses Reinforcement Learning~(RL) while~\cite{related_works:automated_hyperparameter_tuning} uses Bayesian optimization to explore different network architectures. 
\ndss{Specifically,~\cite{related_works:automated_hyperparameter_tuning} has done so through layer-level network morphism and Bayesian optimization as offered by Auto-Keras~\cite{jin2019auto}, the core of their methodology. 
The rationale behind layer-level network morphism is to modify a trained neural network into a new architecture by applying different operations, such as inserting a layer or adding a skip-connection between layers. 
When employing such approaches, attention should be paid as obtained models might overfit~\cite{related_works:automated_hyperparameter_tuning}.
The work in~\cite{related_works:hyperparameter_tuning_RL} is devoted to finding CNNs that are small (in terms of the number of trainable parameters), but exhibit good attack performance; however, choosing the configuration is still (to some extent) guided by the expert through providing a random range of the hyperparameters' values (see Sections~4.2 and 4.5 in~\cite{related_works:hyperparameter_tuning_RL}). } 

InfoNEAT, on the other hand, tunes the \ndss{whole range of parameters (a-d) considered for NAS and additionally, the weights (see Table~\ref{tab:related_NAS} for a summary of the comparison, and Section~\ref{sec:results_discussion} for a comparison between the time complexity of InfoNEAT and existing methods).}
An important feature of NEAT, at the core of InfoNEAT, is that it starts from a small network \ndss{to optimize the networks more efficiently~\cite{stanley2002evolving}. 
With the aid of information-theoretic measures, InfoNEAT is guided to stop without posing any limitation on the number of layers, but when the network is well-trained (Section~\ref{sec:method_info}). 
Moreover, for the purpose of SCA, an adequate fitness function is chosen (see Section~\ref{sec:NEAT_fitness}). 
The stacking technique is also incorporated to scale up the number of classes that NEAT can handle as profiled SCA is a multi-class ML task with a high number of classes (Section~\ref{sec:methodology_training}). 
}

\section{Background and Mathematical Foundations}\label{sec:background}


\subsection{Notations}\label{sec:back_notation}
In this paper, the calligraphic letters, e.g., $\mathcal{X}$, are used to denote sets. 
Moreover, bold letters (e.g., $\textbf{y}$) correspond to matrices and vectors. 
We use the standard notations for mathematical operators defined in the respective sections. 
For a random variable $X$, the corresponding lower-case letter $x$ denotes realizations of $X$. 
Shannon’s definition of MI and CMI is $\textbf{I}(x; y) =\textbf{H}(x) +\textbf{H}(y)- \textbf{H}(x, y)$ and $\textbf{I}(x; y\vert x') =\textbf{H}(x, x')+\textbf{H}(y, x')- \textbf{H}(x, y, x') -\textbf{H}(x')$, where $\textbf{H}$ denotes entropy or joint entropy: $ \textbf{H}=- \sum\limits_{x\in X} p(x)\log p(x)$ and $p(x)$ denotes the probability of random variable $X$ taking value $x$.

\subsection{Profiled Side-channel Analysis}\label{sec:back_profiled}
One of the most powerful forms of side-channel attack is the profiled SCA in which the adversary uses a device that he can control to build a \textit{profiling model} that can then later be used to extract the encryption key from similar devices. 
After the introduction of profiling attacks in~\cite{chari2002template}, due to their close conceptual connection with ML tasks, ML-based attacks were proposed to enhance them. 
Thus, the profiled SCA has two steps: a \textit{profiling} step and an \textit{attack}~(or training and testing as known in ML domain) step. During the first step (profiling), the adversary has access to a test device and can control its (guessable or public) inputs, which are a chunk of plaintext $P$, as well as a part of the cryptographic algorithm's secret key $S$ that the attacker aims to disclose. 
The observations made by feeding these inputs can be seen as an estimation $\hat{\varphi}_s$ of the conditional probability distribution function for every possible $s \in \mathcal{S}$ as defined below cf.~\cite{ascad_paper}. 
$$\varphi_s: (\textbf{x},s) \mapsto \Pr [\textbf{X}=\textbf{x} \; \vert \; (P,S)=(p,s)]. $$
In other words, the traces can be used to estimate $\hat{\varphi}_s$. 
Precisely, for a given set of $\{p_i, s_i\}^n_{i=1}$, the attacker collects $n$ traces $\{x^i_1, x^i_2, \cdots, x^i_k\}^n_{i=1}$, where each trace contains $k$ features ($k\geq 2$). 
Based on this ``profiling dataset'', the adversary constructs a ML model (i.e., a leakage model) that can estimate the probability of inputs for each trace: 
$\hat{\varphi}_{X,P}: (\textbf{x},p) \mapsto \Pr [(P,S) = (p, s)\vert \textbf{X}=\textbf{x}] $ for every $s\in \mathcal{S}$. 

During the second step, the attacker attempts to classify a set of $N_{test}$ traces, the so-called test set corresponding to an unknown $s$, based on the above leakage model.
Formally, the attacker should derive the label for a trace, similar to a ML classification problem: $\textbf{y}=\hat{\varphi}_{X,P}(\textbf{x},p)$, for $\hat{s}\in \mathcal{S}$ so that $\hat{s}=\argmax_{s\in \mathcal{S}}\textbf{y}_k$, where $\textbf{y}_k$ is the $k^{\text{th}}$ entry in the vector $\textbf{y}$. 
For $N_{test}$ traces, a \textit{score} based on the maximum-likelihood of each hypothetical key can be obtained as  $\textbf{d}_k=\prod_{i=1}^{N_{test}}\textbf{y}^i_k$, where $\textbf{y}^i_k$ is the $k^{\text{th}}$ entry in the vector $\textbf{y}^i$ corresponding to the $i^{\text{th}}$ trace. 
Based on this score, the key hypotheses are ranked in a decreasing order based on the rank function (Equation~\eqref{eq:rank}), from which the attacker chooses the key that is ranked first. 
The rank function is defined as follows cf.~\cite{ascad_paper}. 
\begin{equation}\label{eq:rank}
    Rank(\hat{\varphi}, N_{test}) = |\{k \;|\; \textbf{d}_k > \textbf{d}_{k^{*}}\}|,
\end{equation}

\noindent where $k^{*}$ represents the key that has been used during the acquisition of the profiling traces. 
Note that the rank is computed for a collection of $N_{test}$ traces from the test dataset, where $N_{test}$ is increased gradually until the rank is minimized (lower the rank, higher the score).
Since the rank is dependent on the traces used, it is common practice to compute the rank over different chunks of datasets and compute an \textit{average rank}, also called the \textit{guessing entropy}~\cite{maghrebi2016breaking}. 
We use the term ``average rank'' throughout this paper. 

\vspace{0.25ex}

\ndss{\noindent\textbf{Realistic attacker: }
A successful and effective attack results in the average rank that equals zero, whereas the guessing entropy staying at $r$ after mounting the attack means that the attacker must brute force $2^r$ different keys to recover the key~\cite{bhasin2020mind}. 
In more realistic scenarios, it is suggested that the attacker is computationally powerful; however, instead of simply breaking the target by achieving the average rank equal to 0, the attacker attempts to break the target with a minimal number of traces~\cite{related_works:least_num_traces}. 
An example of this is the attacker's power measured as the minimum number of traces tried by the attacker to reach the average rank $<20$ given a model trained using profiling traces~\cite{related_works:SCA_metrics}.  
}

\subsection{Datasets}\label{sec:back_dataset}
\subsubsection{ASCAD Dataset}\label{sec:back_ascad}
ASCAD is a dataset introduced in~\cite{ascad_paper} which consists of electromagnetic~(EM) radiations emitted from the software implementation of AES-$128$ protected with first-order Boolean masking and running on an $8$-bit AVR microcontroller ATMega8515. 
This dataset is structured similar to a typical ML dataset and consists of training (or profiling) traces and testing (or attack) traces.
\majorrevision{
There are two different versions of this dataset which we will name ASCAD\SP{fixed} and ASCAD\SP{var}. ASCAD\SP{fixed} dataset contains 50,000 profiling traces and 10,000 attack traces where each trace is acquired using the same fixed key and contains 700 sample points or features that represent various intermediate values related to the processing of the third S-box. ASCAD\SP{var} on the other hand, contains 200,000 profiling traces and 100,000 attack traces. Here, one out of three profiling traces is acquired using a fixed key, while the rest are acquired using a random key. All attack traces are acquired using a fixed key as well. Each trace in the ASCAD\SP{var} dataset consists of 1400 features.
In addition, both of the datasets contain the \textit{labeled} data, which represents the output of the third S-box during the first round as labels for each trace, leading to $256$ classes. 
Note that throughout this paper, we consider this leakage and the \emph{ID leakage model}, i.e,  the leakage in the form of an intermediate value of the cipher leading to 256 classes; hence, to compare our results with SOTA approaches, we also take into account their results for ID leakage model. 

Both of ASCAD\SP{fixed} and ASCAD\SP{var} datasets are publicly available~\cite{ASCAD_dataset} along with traces desynchronized by  $50$~and~$100$~samples window maximum jitter, which we will refer to as \textbf{\textit{ASCAD\SPSB{fixed}{desync50}}}, \textbf{\textit{ASCAD\SPSB{fixed}{desync100}}}, \textbf{\textit{ASCAD\SPSB{var}{desync50}}}, and \textbf{\textit{ASCAD\SPSB{var}{desync100}}}
}.

\majorrevision{
\subsubsection{AES\_HD Dataset}\label{sec:back_aeshd}
This dataset consists of EM traces acquired from an unprotected implementation of AES-128 on a Xilinx Virtec-5 FPGA with each encryption taking a total of 11 cycles~\cite{related_works:SCA_metrics,kim2019make}.  
A total of 500,000 traces were captured with each trace consisting of 1,250 features. 
The dataset has been extended and made available~\cite{AES_HD_Ext}, which is used in this paper. 
The labels for each of the trace were generated using a HD based leakage model targeting the last round of AES encryption\footnote{Although there are dissimilarity between the leakage model, round key and label calculation in relevant studies, i.e.,~\cite{zaid2020methodology,Wouters_Arribas_Gierlichs_Preneel_2020,related_works:SCA_metrics}, for the sake of comparison, the results as presented in their papers are considered here.}. 
In this paper, we use 45,000 traces for profiling and 4,500 traces for testing and attack purposes.
}

\subsection{Matrix-based R\'enyi Entropy}\label{sec:back_renyi}
One of the recent and major breakthroughs in information theory is the new estimator of R\'enyi entropy in a matrix form as defined below. 
In particular, using this estimator, it has become feasible to understand the information flow without knowing the probability density functions~\cite{yu2019simple,yu2020understanding}. 
This estimator is the basis of the NN selection and algorithm stopping criteria introduced in InfoNEAT. 
\begin{Def}\label{def:back_reney}
(cf.~\cite{yu2019simple}) Given a set of $n$ samples $\{x^i_1, x^i_2, \cdots, x^i_k\}^n_{i=1}$, where each sample contains $k$ ($k\geq 2$) \majorrevision{measurements}, we define the kernels $\kappa_1: \mathcal{X}_1\times \mathcal{X}_1 \mapsto \mathbb{R}$, $\cdots$, $\kappa_k: \mathcal{X}_k\times \mathcal{X}_k \mapsto \mathbb{R}$ ($X_z =x^1_z , \cdots, x^n_z$ with $1 \leq z \leq k$), which are real-valued positive definite and infinitely divisible~\cite{bhatia2006infinitely,yu2019simple}. 
The R\'enyi's $\alpha$-order joint-entropy among $k$ variables is
\begin{align}\label{eq:back_reney_hadamard}
\textbf{J}_\alpha (\textbf{X}_1, \cdots, \textbf{X}_k)=\textbf{S}_\alpha \left(\frac{\textbf{A}_1 \odot \cdots \odot\textbf{A}_{k}}{\text{tr}(\textbf{A}_1 \odot \cdots \odot \textbf{A}_{k})}\right),
\end{align}

\noindent where $(\textbf{A}_1)ij = \kappa_1(x^i_1, x^j_1)$, $\cdots$, $(\textbf{A}_k)ij = \kappa_k(x^i_k, x^j_k)$. 
$\text{tr}(\cdot)$ and $\odot$ denote the transpose and Hadamard product operators, respectively. 
Furthermore, the function $\textbf{S}_\alpha (\cdot)$ is defined as follows.
\begin{align}\label{eq:back_reney}
\textbf{S}_\alpha (\textbf{A})= \frac{1}{1-\alpha}\log_2 (\text{tr}(\textbf{A}^\alpha))=\frac{1}{1-\alpha}\log_2 \sum\limits_{i=1}^n \lambda_i(\textbf{A}^\alpha), \numberthis
\end{align}

\noindent where $\lambda_i(\textbf{A})$ denotes the $i^{\text{th}}$ eigenvalue of $\textbf{A}$. 

\end{Def}
There is a relationship between the matrix $\textbf{A}$ as defined in Equation~\eqref{eq:back_reney} and the Gram matrix $\textbf{K}$. 
In this regard, $\textbf{A}_{ij} = \frac{\textbf{K}_{ij}}{n\sqrt{\textbf{K}_{ii}\textbf{K}_{jj}}}$. 
Moreover, the information quantities are estimated using the above mentioned matrix-based R\'enyi entropy with with $\alpha = 1.01$ to approximate Shannon's entropy as suggested in~\cite{giraldo2013rate,giraldo2014measures}. 

\section{NeuroEvolution of Augmenting Topologies}\label{sec:neat}

\begin{figure}[t]
\centering
\includegraphics[width=0.7\columnwidth]{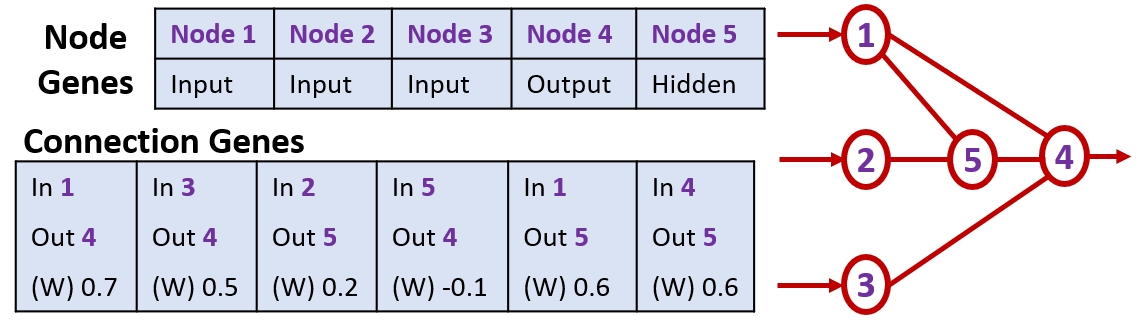} 
\caption{A typical genome or network created in NEAT. $(W)$ denotes the weight assigned to each connection.}
\label{fig:genome_network}
\end{figure}

\begin{figure}[t]
\centering
\includegraphics[width=0.7\columnwidth]{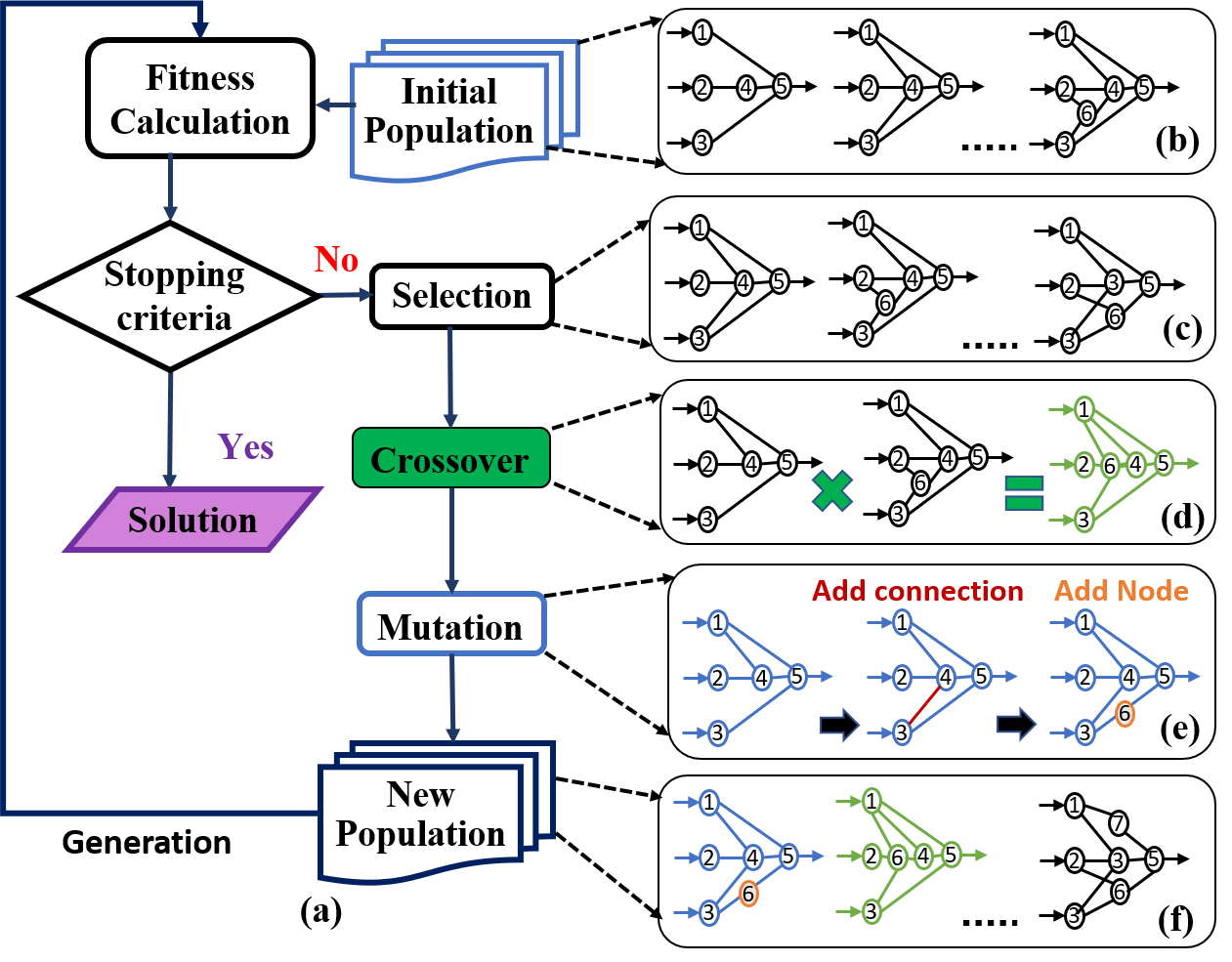} 
\caption{A general overview of the evolution mechanism involved in NEAT, and consequently, InfoNEAT. 
\ndss{Note the multi branches and skip connections that have been used in step (e). }
\majorrevision{Note that the population depicted here includes only one species.} }
\label{fig:neat_generations}
\end{figure}


This section gives an overview of how the neuroevolution of augmenting topologies (NEAT) algorithm works as part of InfoNEAT. 
For more details, the reader is referred to~\cite{stanley2002evolving}. 
The fundamental process taking place in NEAT is similar to the evolution of organism's genomes, where genomes represent NNs, see Figure~\ref{fig:genome_network}. 
In particular, the main idea behind NEAT is the concept of neuroevolution, which searches for network
topologies by means of evolutionary algorithms (EAs). 
In doing so, the EA optimizes hyperparameters of deep
networks, specifically, the weights of individual neurons and their inter-connections are evolved in each step, called a \textit{generation}, see Figure~\ref{fig:neat_generations}(a). 
\ndss{In the context of SCA, the number of generations is equivalent to the minimum number of attempts made by the attacker to tune the hyperparameters, cf.~\cite{related_works:least_num_traces}.  }

Although NEAT variants can be comprised of various types of NNs, e.g., convolutional NNs, we stick to a variant evolving MLP-like NNs in this paper. 
\majorrevision{In fact, the configuration of NNs (so-called genomes) is different from typical multi-layer perceptrons (MLPs), where each hidden neuron is connected to every neuron in the previous and next layers. 
This irregular topology (compared to the structure of MLPs) discovered by NEAT is one of the most important aspects of NNs and contributes to their special behavior. 
As explained in, e.g.,~\cite{stanley2002evolving}, the selection of topology throughout evolution as performed by NEAT brings various advantages, including (1) saving time spent on selecting the topology manually, (2) faster learning speed due to topology evolution, i.e., minimizing and growing it incrementally, and (3) along with the evolution of weights, it significantly enhances the performance of the NNs. }

NNs generated by NEAT are grouped in species based on their randomly chosen structures. 
In each generation, several species (i.e., sub-populations) exist that share a similar topology (i.e., the way that the nodes are connected to each other). 
In this regard, the genomes in different species can have different sizes. 
Independent from other species, each species evolves proportionally to its fitness so that the weights of NNs are learned. Learning the weights of NNs has conventionally been performed without backpropagation. 
Figure~\ref{fig:neat_generations} illustrates the main steps involved in NEAT (see Algorithm~\ref{alg:NEAT}) as explained below. 

\vspace{5pt}\noindent\textbf{Initial population (line \#1 in Algorithm~\ref{alg:NEAT}): }
The population considered by NEAT contains NNs, i.e., the genomes (see Figure~\ref{fig:genome_network}) that are denoted by $g^{j}_{t,i}$, where superscript $j$ ($1\leq j \leq \ell$) indicates the $j^{\text{th}}$ layer. 
Moreover, the subscripts $t$ and $i$ denote the generation and the $i^{\text{th}}$ NN in the species, respectively. 
In the first generation $t=1$, the total number of NNs in all species is $N$ and each species contains $N_{s,t=1}$ NNs. 
For the sake of simplicity and without loss of generality, we explain the learning/evolution phase for one species. 
Features at the input of each genome and the output class labels are denoted by $\textbf{X}$ and $\textbf{Y}$, respectively. 
The evolution process begins with a set of NNs with minimal complexity. 
In the first generation, in a species, each NN $g^{1}_{1,i}$ ($1 \leq i \leq N_{s,1}$) is composed of $j=1$ hidden layer. 
The output of each layer is a function of the input $\textbf{X}$, the matrix of weights between $j^{\text{th}}$ and $(j + 1)^{\text{th}}$ layers $\textbf{W}_j$ and the vector of biases for the $(j + 1)^{\text{th}}$ layer $\textbf{b}_{j+1}$ ($1\leq j \leq \ell$). 
More specifically, for a set of weights and biases $\theta_j=\{\textbf{W}_j, \textbf{b}_{j+1}\}$, the function $f: \mathbb{R}^{\vert \theta_j\vert} \rightarrow \mathbb{R}^{\vert \theta_{j+1}\vert}$ is applied to the weights, biases and inputs to generate the output at the $(j + 1)^{\text{th}}$ as follows: 
$\hat{\textbf{y}}_{j+1}= f (\hat{\textbf{y}}_{j}\textbf{W}_j + \textbf{b}_{j+1}),$
where $\hat{\textbf{y}}$ and $f(\cdot)$ denote the estimated label and the activation function, respectively. 

\vspace{5pt}\noindent\textbf{Fitness evaluation and selection (lines \#3--5 in Algorithm~\ref{alg:NEAT}): }
One of the differences between the NNs usually applied in various domains and NEAT as a neuroevolutionary algorithm is that all the NNs in a species are evaluated based on a fitness function. 
According to this probabilistic evaluation, the algorithm decides which NNs would be successful in the next generations. 
This is accomplished by assigning a fitness value to each genome. 
Afterward, the fitness values are taken into account by a selection operator, e.g., tournament selection~\cite{galvan2021neuroevolution} \majorrevision{that examines a small random subset of the population to find the fittest individual} (see Figure~\ref{fig:neat_generations}(c)). 

\vspace{5pt}\noindent\textbf{Crossover (lines \#6 in Algorithm~\ref{alg:NEAT}): }
In the next step, crossover operation, also known as recombination, is performed, where the genetic information (e.g., the parameters of the NNs) of two selected individuals are combined as shown in Figure~\ref{fig:neat_generations}(d). 
\majorrevision{For this, genomic distance is considered to define species within which the combination happens. 
This distance is a combination of the number of disjoint genes or nodes (i.e., derived from different ancestors), as well as the average weight differences of matching genes or nodes between two networks~\cite{stanley2002evolving}.
Within a species, the crossover is based on the historical marking, used to label each connection based on their ancestry. For crossover between two parents, their connection genes are lined up according to their historical markings. The genes that have the same markings are swapped at random while the remaining genes or disjoint genes are stacked at the end to create a new offspring, cf.~\cite{stanley2002evolving}. }

\vspace{5pt}\noindent\textbf{Mutation (lines \#7 in Algorithm~\ref{alg:NEAT}): }
For each non-crossover individual, random changes are made into the NNs based on the mutation operator, as shown in Figure~\ref{fig:neat_generations}(e).  
Specifically, nodes and connections are added incrementally to these NNs in each generation to update configurations and parameters. The steps above are repeated until a condition is met, e.g., typically when a maximum number of generations are executed. 

{\SetAlgoNoLine
\begin{algorithm}[t]
\small
 \caption{NEAT Algorithm cf.~\cite{galvan2021neuroevolution}}\label{alg:NEAT}
\setstretch{0.9}
\KwIn{Batch size data of input dataset $D$; Population size $N$\;
    Initial number of hidden nodes $n_h$; Fitness threshold \majorrevision{$L_{TH}$}\;
    Connection add probability $P_c$; Node add probability $P_n$\;
    Number of generations $T$; Compatibility threshold $t_c$\;
    Weights and bias mutation rate $P_{w}$ and $P_b$, respectively\;
     }
 \textbf{Initialization: }Generate a set of genomes or networks $g_{t=1,i}$ ($1 \leq i \leq N$) randomly based on $N$ and $n_h$\;
 \For{$t=1,2,...,T$}{
   \textcolor{gray} {\textbf{Fitness evaluation: } Compute the fitness (e.g., cross-entropy loss) for $g_{t,i}$\;
  \textbf{if} \textit{fitness values} $L_{t,i}\leq L_{TH}$
  \textbf{then} break;
  \textbf{else} continue; 
  }\texttt{\\}
    \textcolor{gray} {\textbf{Selection: }Select the best individuals and produce a new generation $g_{t+1,i}$ from $g_{t,i}$\;}
    \textbf{Crossover: }Individuals with genomic distance $< t_c$ are part of the same species and are selected for crossover\;
    \textbf{Mutation: }For each individual $g_{t,i}$, the mutation of weights and bias is performed based on $P_w$ and $P_b$ respectively and the structural mutation is performed based on $P_c$ and $P_n$\;
\textbf{end}
 }
\end{algorithm}}

\section{InfoNEAT}\label{sec:infoneat}
This section gives details on InfoNEAT, built upon NEAT. 
Variants of the NEAT algorithm have been developed accordingly, which are used for various tasks, including regression~\cite{hagg2017evolving}, classification~\cite{hagg2017evolving,stanley2002evolving}, and reinforcement learning (RL)~\cite{schrum2015discovering}. Nevertheless, in this paper, we stick to NEAT's application to classification.
First, we explain which fitness function can be applied to tailor NEAT for SCA. 
It is followed by how the information-theoretic stopping criteria are defined for InfoNEAT. 
Finally, we discuss the challenges that were faced when designing and implementing InfoNEAT. 

\subsection{Adequate Fitness Function for SCA }\label{sec:NEAT_fitness}
The fitness function evaluating the performance of the NNs in each generation helps to select the NNs that will be evolved in the next generations. 
For a given genome (e.g., $i^{\text{th}}$ NN) in the $t^{\text{th}}$ generation, InfoNEAT employs the categorical cross-entropy loss function to compute the loss $L_{t,i}$ formulated as
$$ L_{t,i}=-\sum\limits_{k=1}^{\vert \textbf{y}^\ell_{t,i}\vert} y_k\log \hat{y}_k,$$
where $y_k$ and $\hat{y}_k$ are the $k^{\text{th}}$ entries in $\textbf{y}^\ell_{t,i}$ and $\hat{\textbf{y}}^\ell_{t,i}$, respectively. 
When it comes to computing the cross-entropy for classification tasks, the terms ``cross-entropy'' and ``negative log-likelihood'' are used interchangeably~\cite{murphy2012machine}. 
This is of great importance to SCA as it has been proven that negative log-likelihood (NLL) is inversely related to ``perceived information''~\cite{renauld2011formal,masure2020comprehensive,bronchain2019leakage}. 
The latter refers to the generalization of the mutual information between the side-channel traces and the leakage profiling model (i.e., the ML trained on the traces). 
In other words, the perceived information quantifies how well the ML model is trained. 
More interestingly, minimizing the NLL loss function (similarly, cross-entropy) during NN training is asymptotically equivalent to maximizing the perceived information and improving the trained NN performance cf.~\cite{masure2020comprehensive}. 


\subsection{Information Theoretic Criteria}\label{sec:method_info}
Conventionally, a maximum number of generations is defined, usually accompanied by another stopping criterion relying on the heuristic, e.g., if the accuracy of the best NN (i.e., the NN with the highest accuracy) does not improve after some generations, halt the evolution. 
To address this, InfoNEAT applies information theory-based approaches to select the genomes (NNs) to be evolved (see the part highlighted in gray in Algorithm~\ref{alg:NEAT}) as well as to stop the evolution process. 
Although InfoNEAT shares a similarity with~\cite{related_works:MI_SCA}, namely relying on an information-theoretic measure to stop the training, we neither train NNs by minimizing the information bottleneck (IB) function~\cite{amjad2019learning}, nor consider the so-called fitting and compression phases~\cite{saxe2019information} as applied in~\cite{related_works:MI_SCA}; hence, the issues inherent to these methods~\cite{saxe2019information} are avoided by InfoNEAT. 


In our approach, each mutated NN is seen as a randomly permuted one. Intuitively, making any change in a NN (i.e., evolving the NN in the next generation) should result in a monotonic decrease in the conditional mutual information (CMI) that is $\textbf{I}\left( \hat{\textbf{y}}^j_{t+1,i};\textbf{y}\vert \hat{\textbf{y}}^j_{t,i} \right)$, where $\hat{\textbf{y}}^j_{t,i}$ denotes the output of a given layer in $g^j_{t,i}$. 
This is according to permutation tests~\cite{good2013permutation} and well-described in~\cite{franccois2007resampling} as an impact of adding useless variables (i.e., parameters associated with evolved NN). 
Below, we formalize this more precisely.  

{\SetAlgoNoLine
\begin{algorithm}[t]
\small
\setstretch{0.1}
\KwIn{Number of genomes within a species $N_s$; All the genomes in a species $g^{j}_{t,i}$ ($1 \leq i \leq N_s$ and $1 \leq j \leq \ell$)\;}
\KwResult{Best genome within a species $g^{j}_{t,*}$\;}
\textbf{Initialization:} Calculate the loss values $L_{t,i}$ for $1 \leq i \leq N_{s,t}$\;
  \eIf{number of genomes with $\min\limits_{1 \leq i \leq N_s}(L_{t,i})$ is one}{
   break\;
   }{
   $G = \{g_{t,i}\vert\: L_{t}=\min\limits_{1 \leq i \leq N_{s,t}}(L_{t,i})\}$ \; 
   \For{$g_{t,i}\in G$}{
    \For{$j = \ell, \ell-1, \cdots, 1$}{
    $\text{CMI}=\textbf{I}\left(\hat{\textbf{y}}^j_{t+1,i};\textbf{y}\vert \hat{\textbf{y}}^j_{t,i} \right)$\;
    \eIf{number of genomes with $\min\limits_{1 \leq i \leq N_{s,t}}(\text{CMI})$ is one}{\majorrevision{Output $g_{t,i}$ and} break\;}
    {continue\;}
    }}}
 \caption{Genome selection based on fitness and CMI}\label{alg:selection_criteria}
\end{algorithm}}

\subsubsection{Selection of the Best Genomes}\label{sec:method_genome_selection}
The selection criterion offered in NEAT is based on the loss function as explained in Algorithm~\ref{alg:selection_criteria}, line \#2. 
\newversion{However, if the genomes perform almost similarly well as can happen in the last generations, loss-based selection criteria would not be effective. }
For this purpose, we combine this with CMI-based criterion explained below.

As discussed before, from each species, a set of NNs are evolved in the next generation. 
The evolution of NNs from one generation to the next is considered as a permutation (without permuting the corresponding $\textbf{y}$). 
\newversion{Precisely, when a genome is evolved to another one in the next generation, the mutual information between the output of the $(j)^{\text{th}}$ layer and the output class labels in generation $t$ and $t+1$ are $\textbf{I}\left( \hat{\textbf{y}}^j_{t,i};\textbf{y}\right)$ and $\textbf{I}\left( \hat{\textbf{y}}^j_{t+1,i};\textbf{y}\right)$, respectively. 
The difference between these two is the CMI 
$\textbf{I}\left( \hat{\textbf{y}}^j_{t+1,i};\textbf{y}\vert \hat{\textbf{y}}^j_{t,i} \right)$, which is non negative and rarely equals zero in practice because of the statistical variation~\cite{cover1999elements,vinh2014reconsidering}.  
Hence, from one generation to the next, making an unnecessary change to the well-trained genome (almost surely) increases the mutual information between $\hat{\textbf{y}}^j$ and $\textbf{y}$ that is against the training goal. The higher this increase is, the less useful the change made to the genome would be. }

With regard to this principal, to choose one or more NNs from a species, InfoNEAT discards NNs (e.g., $(i+1)^{\text{th}}$ genome), when the CMI $\textbf{I}\left( \hat{\textbf{y}}^j_{t+1,i+1};\textbf{y}\vert \hat{\textbf{y}}^j_{t,i+1} \right)$ is not significantly smaller than $\textbf{I}\left( \hat{\textbf{y}}^j_{t+1,i};\textbf{y}\vert \hat{\textbf{y}}^j_{t,i} \right)$ (see lines \#6--15 in Algorithm~\ref{alg:selection_criteria}). 
This means that if the mutated NNs cannot outperform the respective ones in the previous generation, those NNs are discarded. 
\newversion{To implement this CMI-based criterion, however, technical difficulties regarding the computation of the CMI should be resolved as discussed below. 
}

\vspace{0.25ex}\noindent\textbf{CMI computation based on Matrix-based R\'enyi's $\alpha$-entropy: }
This notion encompasses the extension of Shannon's entropy; however, in its traditional form, the probability distribution function (PDF) should be accurately estimated. 
To cope with this, we follow the procedure presented in~\cite{yu2019multivariate,giraldo2014measures} that relies on the principle of the Gram matrix obtained from evaluations of a positive definite kernel from data samples, see Definition~\ref{def:back_reney}. 
This allows a direct estimation of the entropy and joint entropy between two or multiple variables from data without PDF estimation. 
The multivariate matrix-based R\'enyi's $\alpha$-entropy can be applied to estimate the CMI in high-dimensional space as follows cf.~\cite{yu2019simple} (see Equation~\eqref{eq:back_reney_hadamard}). 
Suppose that $\vert \hat{\textbf{y}}^j_{t+1,i}\vert=k_{t+1}$ and $\vert \hat{\textbf{y}}^j_{t,i}\vert=k_{t}$ denoting the number of outputs at the $j^{\text{th}}$ layer of the $i^{\text{th}}$ genome in the generations $t+1$ and $t$, respectively. Then, the CMI is
\begin{align}\label{eq:renyi}
\textbf{I}_\alpha (\{\textbf{C}_1, \textbf{C}_2, \cdots, \textbf{C}_{k_{t+1}}; \textbf{B}\vert \{\textbf{A}_1, \textbf{A}_2, \cdots, \textbf{A}_{k_{t}}\})= \notag\\
= \textbf{S}_\alpha \left( \frac{\textbf{A}_1 \odot \cdots \odot \textbf{A}_{k_{t}}\odot \textbf{C}_1 \odot \cdots \odot \textbf{C}_{k_{t+1}}}{\text{tr}(\textbf{A}_1 \odot \cdots \odot \textbf{A}_{k_{t}}\odot \textbf{C}_1 \odot \cdots \odot \textbf{C}_{k_{t+1}})} \right)+\notag\\
+\textbf{S}_\alpha \left( \frac{\textbf{A}_1 \odot \cdots \odot \textbf{A}_{k_{t}}\odot \textbf{B} }{\text{tr}(\textbf{A}_1 \odot \cdots \odot \textbf{A}_{k_{t}}\odot \textbf{B}} \right)
- \textbf{S}_\alpha \left( \frac{\textbf{A}_1 \odot \cdots \odot \textbf{A}_{k_{t}}}{\text{tr}(\textbf{A}_1 \odot \cdots \odot \textbf{A}_{k_{t}}} \right)-\notag\\
- \textbf{S}_\alpha \left( \frac{\textbf{A}_1 \odot \cdots \odot \textbf{A}_{k_{t}}\odot \textbf{B} \odot \textbf{C}_1 \odot \cdots \odot \textbf{C}_{k_{t+1}}}{\text{tr}(\textbf{A}_1 \odot \cdots \odot \textbf{A}_{k_{t}}\odot \textbf{B} \odot \textbf{C}_1 \odot \cdots \odot \textbf{C}_{k_{t+1}})} \right).\numberthis
\end{align}
In Equation~\eqref{eq:renyi}, $\textbf{A}_1, \cdots, \textbf{A}_{k_{t}}, \textbf{B} , \textbf{C}_1 , \cdots , \textbf{C}_{k_{t+1}}$ denote the Gram matrices evaluated over $\hat{\textbf{y}}^j_{t,i}$, $\textbf{y}$, and $\hat{\textbf{y}}^j_{t+1,i}$, respectively. 
Moreover, $\textbf{S}_\alpha(\cdot)$ and $\odot$ denote the R\'enyi's $\alpha$-entropy and the Hadamard product (see Definition~\ref{def:back_reney}). 
This equation is the core of Algorithm~\ref{alg:selection_criteria} (line \#8) for how InfoNEAT selects the genomes to be evolved. 
Note that this algorithm is run in every generation when the genomes' parameters are updated. 

As can be seen in Algorithm~\ref{alg:selection_criteria}, the CMI is computed to compare the genomes in two consecutive generations in a layer-wise manner. 
Interestingly, this comparison is first applied to the last layers of the genomes in the sense that if the CMI values are the same for the last layers of the genomes $j=\ell$, the second to last layer $j=\ell-1$ is considered and so on. 
This is due to the fact that the layer-wise mutual information between the labels and the outputs of a layer is minimized at the last hidden layer. 
This value increases so that its maximum can be observed at the first hidden layer~\cite{shwartz2017opening,saxe2019information,zhang2016understanding}.  
Therefore, our layer-wise comparison implies that for the best genome selected through Algorithm~\ref{alg:selection_criteria}, the mutual information between the labels and the outputs of the last layer can stay minimized~\cite{fleuret2004fast}. 


\subsubsection{Stopping Criteria}\label{sec:method_stopping}
Our method to deal with the definition of a stopping criterion relies on the notions discussed for selecting the best genome from a species. 
In fact, the stopping mechanism can be seen as a continuation of the process associated with best genome selection. 
Compared to one of the most relevant approaches presented in~\cite{yu2019simple}, no threshold is needed to stop the algorithm. 
InfoNEAT makes a decision based on the change in the CMI value, following the so-called CMI-permutation concept. 
The permuted NNs are evolved by InfoNEAT automatically and randomly evolved from one generation to the next.  
To halt the process, we monitor not only the CMI values, but also the fitness values that are the cross-entropy loss (see Section~\ref{sec:neat}).  

{\SetAlgoNoLine
\begin{algorithm}[t]
\small
\setstretch{0.1}
\KwIn{Best genomes in the current generation $g^{\ell}_{t,*}$ and in the generation $g^{\ell}_{t-1,*}$ (i.e., the output of Algorithm~\ref{alg:selection_criteria});
Genomes evolved from $g^{\ell}_{t,*}$ in the next generation $g^{\ell}_{t+1,i}$ ($1 \leq i \leq N_{s,t+1}$)\;}
\vspace{5pt}
\KwResult{Stop the training or evolution process and deliver the best genome $g^{\ell}_{T,*}$ ($T\geq 2$)\;}
\textbf{Initialization: }Calculate the loss $L_{t+1,i}$ for $g^{\ell}_{t+1,i}$ as well as $L_{t,*}$, i.e., the loss for the best genome $g^{\ell}_{t,*}$\;
$\text{CMI}_{t}=\textbf{I}\left(\hat{\textbf{y}}^\ell_{t,*};\textbf{y}\vert \hat{\textbf{y}}^\ell_{t-1,*} \right)$\;
\For{$i=1,\cdots,N_{s,t+1}$}{
\eIf{$L_{t,*} < L_{t+1,i}$}{break and $g^{\ell}_{T,*}=g^{\ell}_{t,*}$\;}
    {
$\text{CMI}_{t+1}=\textbf{I}\left(\hat{\textbf{y}}^\ell_{t+1,i};\textbf{y}\vert \hat{\textbf{y}}^\ell_{t,*} \right)$\;
\eIf{$\text{CMI}_{t} < \text{CMI}_{t+1}$}{break and $g^{\ell}_{T,*}=g^{\ell}_{t,*}$\;}
{continue\;}
}
}
 \caption{Stopping criteria based on fitness and CMI}\label{alg:stopping_criteria}
\end{algorithm}}

Algorithm~\ref{alg:stopping_criteria} explains this further. 
In the initialization phase (line~\#1), the loss function for the genomes in the current generation $t+1$ and the best genome $g^{\ell}_{t-1,*}$ delivered by the Algorithm~\ref{alg:selection_criteria} is computed. 
An interesting observation is that to define the stopping criteria, similar to Algorithm~\ref{alg:selection_criteria}, the last layers of the genomes are taken into account. 
This is due to the fact that the last hidden layer carries the highest level of mutual information between the output of hidden layers and the labels~\cite{zhang2016understanding,shwartz2017opening}. 
In this regard, and according to the CMI-permutation principle, the stopping criteria encompasses the CMI between the output of the last hidden layer and the labels conditioned on the best genome created in the previous generations (lines \#2,7 in Algorithm~\ref{alg:stopping_criteria}). 
Moreover, the degradation in the loss (i.e., an increase in the cross-entropy loss) is considered as in the line \#4 in Algorithm~\ref{alg:stopping_criteria}. 
If the degradation is not observed, the CMI values are compared to ensure that the permuted genomes, i.e., the NNs in generation $t+1$, outperform their respective descendant $L_{t,*}$. 
If not, the algorithm halts and outputs $L_{T,*}=L_{t,*}$ that is the best genome at the last generation denoted by $T$. 
\textit{Note that no stopping criterion has been previously defined for the NEAT algorithm, as shown in Algorithm~\ref{alg:NEAT}.}


\subsection{Training and Testing Phases}
\label{sec:methodology_training}
\begin{figure}[t]
\centering
\includegraphics[width=0.8\columnwidth]{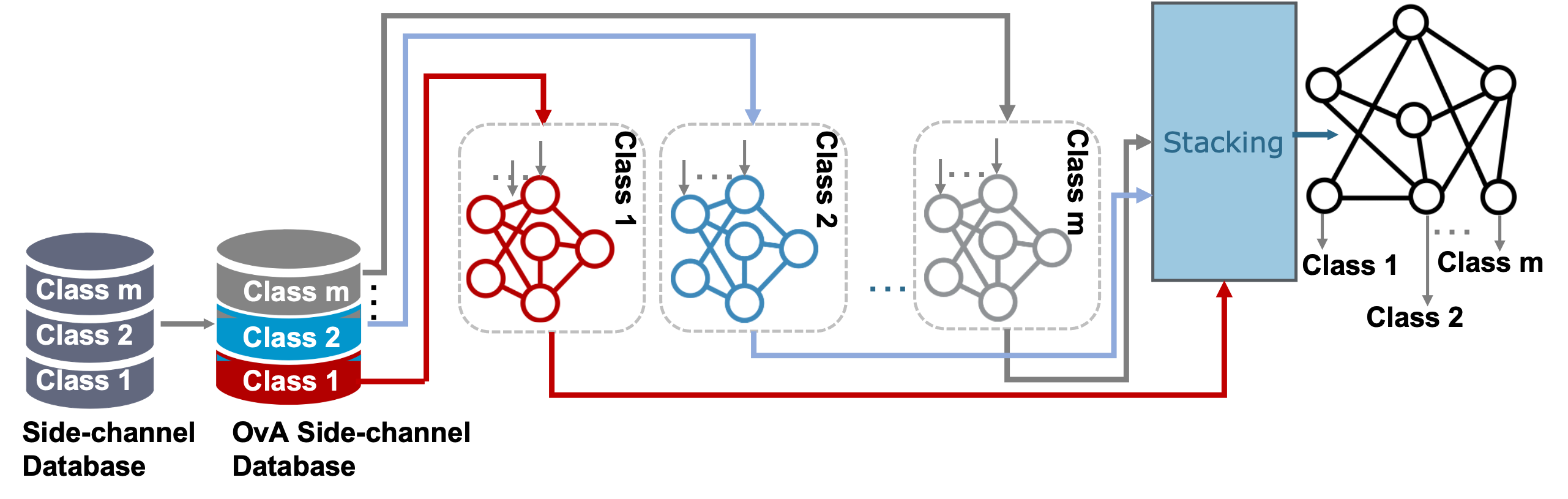} 
\caption{Schematic of the NEAT algorithm configured to deliver a stacked model. 
For each class, a sub-model is trained by feeding batches from One-vs-All (OvA) database. }
\label{fig:neat_stacking}
\end{figure}

The training phase of InfoNEAT involves running the algorithms corresponding to evolution process enhanced by the best genome or network selection, and training stopping, i.e., Algorithms~\ref{alg:NEAT}-\ref{alg:stopping_criteria}. 
Neuroevolution techniques in Algorithms~\ref{alg:NEAT} evolve the weights and structure of a network from a simple starting point, and usually develop a minimal and a generalizable NN. 
However, it is challenging for the NEAT to optimize all parts of the network if the learning task concerns multi-class classification~\cite{mcdonnell2018divide}. 
To tackle this, InfoNEAT uses the One-vs-All (OvA) classification technique to develop $m$ sub-models (i.e., NNs or genomes) corresponding to the number of labels or classes (i.e., 256 sub-keys).
\majorrevision{
This is similar to what has been proposed in the context of neuroevolution in~\cite{mcdonnell2018divide}.
However, the methods for obtaining the final classification result (i.e., the class an unseen trace belongs to) are different. 
The approach employed in~\cite{mcdonnell2018divide} simply selects the class, whose corresponding sub-models outputs the highest probability. 
\ndss{This is not helpful when considering SCA with a larger number of classes (256 vs. 26 considered in~\cite{mcdonnell2018divide}) making the classification more prone to noise~\cite{related_works:SCA_metrics}. 
Under such a complicated scenario, soft classifiers, e.g., logistic regression, work better~\cite{liu2011hard}. }
Hence, InfoNEAT takes advantage of the stacking method, where a meta-learner (i.e., a classifier at the second layer, see Figure~\ref{fig:neat_stacking}) is trained to combine the output of the sub-models so that the final, stacked model outperforms each sub-model. }
Notice the difference between stacking of ML models trained for a multi-class task and our approach, where a set of $m$ OvA models are stacked to combine them and further improve the predictive performance. 
\majorrevision{For this, we apply logistic regression, trained to best combine the predictions from each of the sub-models. 
This is, of course, advantageous as the class label prediction can be improved through training not only the sub-models, but also the meta-learner (as an example, compare the green and orange curves in \minr{Figure~\ref{fig:results_stacking}}). }

\ndss{\noindent\textbf{OvA vs. ensemble learning proposed in~\cite{related_works:ensemble_SCA}:}
Although the bagging ensemble technique applied in~\cite{related_works:ensemble_SCA} and our stacking method share some similarities, i.e., being from the same ensemble learning category of ML algorithms, there is a crucial difference between them. 
InfoNEAT combines $m=256$ sub-models, corresponding to sub-keys, and deliver \emph{one} model (i.e., stacked model, here denoted by $M$). 
\cite{related_works:ensemble_SCA} has suggested combining 50 models trained on \emph{all sub-keys}, i.e., $M_1, \cdots,M_{50}$, through the bagging technique; hence, their proposed technique is different in nature from ours.}


\vspace{0.25ex}\noindent{\textbf{Training the sub-models:}}
\label{sec:training_sub_models}
We train the sub-models by running InfoNEAT algorithm $m$ times using $m$ different, respective sub-datasets, see Figure~\ref{fig:neat_stacking}. 
If $k$-fold cross-validation is used, each of these $m$ different datasets contains $k$ folds. For each label or class, we make sure that the dataset contains an equal number of traces belonging and not belonging to that particular class as required by the OvA method. The labels in these datasets are then modified accordingly using one-hot encoding where the length of each label is $m$. If the data belongs to the particular class, only the index corresponding to that class is $1$. 
All the indices in the labels for traces not belonging to the class are set to $0$. 

\vspace{0.25ex}\noindent{\textbf{Training the stacked model:}}
\label{sec:training_stack}
After $m$~number of sub-models are trained, a stacked model is trained which basically combines the predictions from all the sub-models and outputs a final prediction as shown in Figure~\ref{fig:neat_stacking}. 
To train the stacked model (meta-learner), a dataset is prepared by involving the predictions from all the sub-models. 

\vspace{0.25ex}\noindent{\textbf{Testing phase:}}
The trained meta-learner is used against the test dataset to return a set of predictions (or \textbf{average ranks} in the case of SCA). 
Notice that we do not use the common ML metrics to test our model due to the fact that the ML metrics, e.g., the accuracy, do not provide relevant information to the attackers; hence, it is not guaranteed that a model with good performance based on ML metrics necessary performs well in the case of SCA~\cite{related_works:SCA_metrics}.
If $k$-fold cross-validation is applied, the training/testing process is repeated $k$ times, and the average of the average ranks is reported. 
\subsection{Discussion of Practical Implementation}
Various practical challenges were encountered and resolved during InfoNEAT's design and execution. They are itemized and discussed below along with pointers to our results.


\vspace{0.25ex}\noindent{\textbf{Bias and variance in NNs:}}
A common problem with stochastic algorithms involved in training the NNs is the high variance in the network similar to the case of SCA~\cite{van2019bias}. In other words, every time the model is fit to a new data point, the network has different sets of weights and parameters which in turn makes different predictions. 
To deal with this, the weight initialization method is chosen carefully for InfoNEAT (see Appendix~B) and the evolution is further controlled by monitoring the fitness and CMI values.  
On top of this, stacking is used that has also been shown to produce models with lower bias compared to the sub-models used. 
Furthermore, for datasets with fixed key, we employ the k-fold cross-validation technique to reduce the variability of the stacked model associated with learning new data points (see Figure~\ref{fig:results_learning_curve} in Appendix~B showing an example of our sub-models that is neither underfitting nor overfitting). 


\vspace{0.25ex}\noindent{\textbf{Search space explored by NEAT:}}
The search space in the case of NNs is the set of all possible weight configurations -- the wider and deeper the network is, the larger the search space is. In that sense, the architecture of the network (i.e., the number of layers and number of hidden nodes) can also be considered as part of the search space since they affect the size of the network. Neuroevolution techniques such as NEAT evolve both the weights and NN architecture through crossover and mutation~\cite{galvan2021neuroevolution}. 
Thus, compared to traditional optimizers (such as the stochastic gradient descent, SGD) which only optimize the weight, the search space for neuroevolution techniques is considerably larger~\cite{aly2019optimizing}. In InfoNEAT, this challenge is simplified. InfoNEAT starts the evolution process from small sized network and uses the evolutionary operators, namely the mutation and crossover operators, to explore the search space. But unlike NEAT, InfoNEAT uses the CMI values rather than just the fitness values to help guide evolution. 
This also helps to achieve an optimally sized network as the hidden nodes are only added if the CMI values do not decrease monotonically. 

\begin{figure}[t]
\centering
\includegraphics[width=0.4\columnwidth]{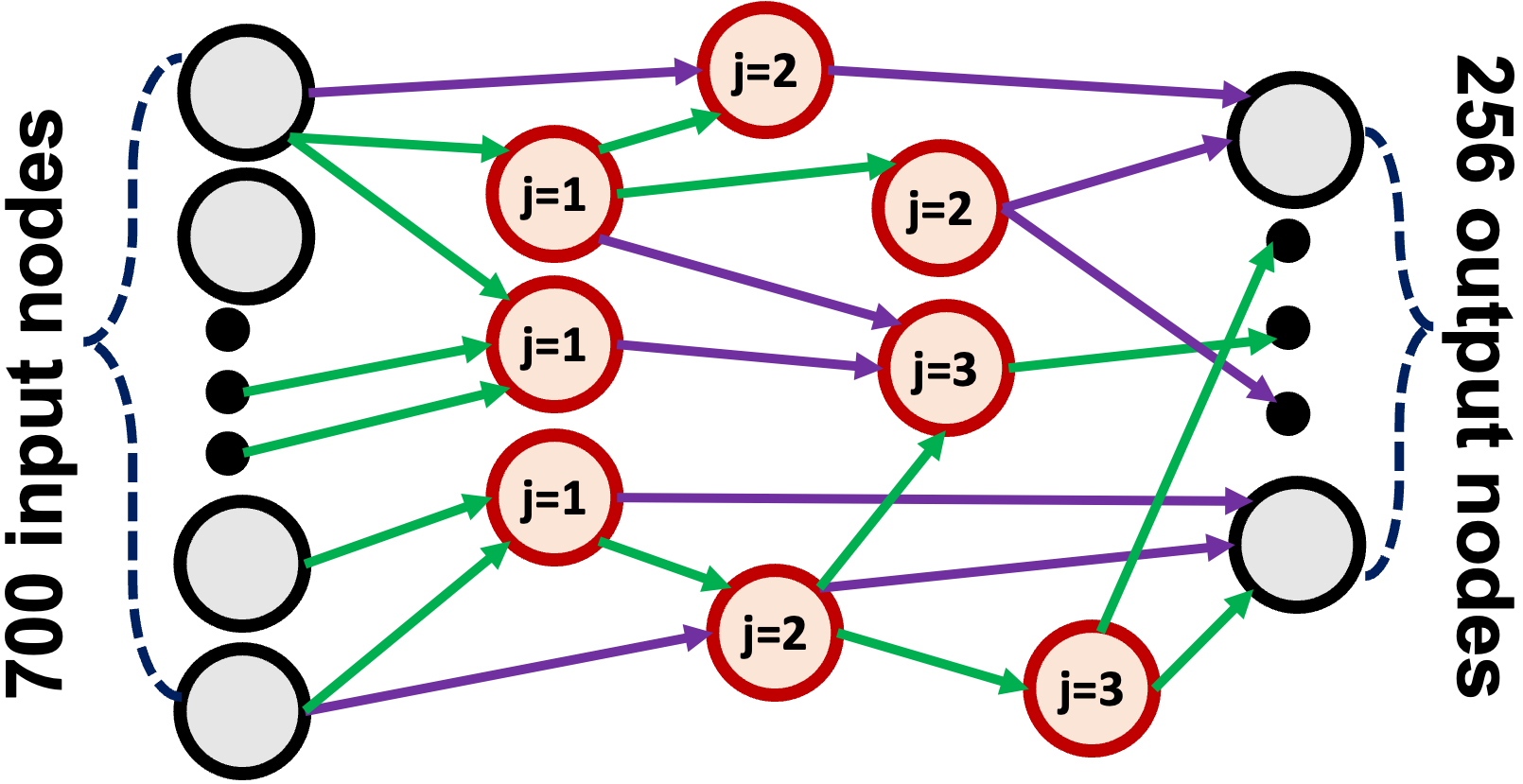} 
\caption{Example of the irregular InfoNEAT architecture used for SCA with \majorrevision{5 layers (3 hidden layers based on the node connections). The nodes in one layer can have a direct connection to a layer other than the immediate next layer (as highlighted in purple), e.g., the top input node has a direct connection to a node in hidden layer $j=2$, but not $j=1$.}}
\label{fig:InfoNEAT_architecture}
\end{figure}

\vspace{0pt}\noindent{\textbf{Evaluation of multiple \textit{distinct} networks:}}
A salient feature of NEAT that makes it preferable over other network optimization methods is that NEAT evaluates a population of different networks all at the same time. However, this is only true if the networks are inherently distinct. One of the NEAT methods that help maintain diversity among networks is \textit{speciation}. Speciation has been typically effective for penalizing similar networks by looking at the fitness and the structure of the network. The speciation parameters thus have to be tuned very carefully because of the associated time complexity of evaluating a typical SCA network and converging towards a solution. We discuss some of these parameters in more detail in Section~\ref{sec:results_neat_architecture}, namely the \textit{compatibility threshold} which helps divide a population of genomes into different species where the genomes between two species are inherently distinct.

\vspace{0pt}\noindent{\textbf{Automatic evaluation of CMI for analyzing and finding the best genome every generation and implementing the stopping criteria:}}
The methods proposed in the literature calculate the CMI for typical CNNs and MLPs to determine the number of filters and select features, respectively~\cite{yu2019simple,yu2020understanding}. However, the NN evolved using NEAT is an unusual network as nodes from multiple layers can have a direct connection to each other as opposed to a typical MLP network, see Figure~\ref{fig:InfoNEAT_architecture}. We have addressed this by calculating the CMI for each node based on all of its connections.


\section{Results}\label{sec:results}

\subsection{Experimental Setup}
\label{sec:results_experimental_setup}
Experiments in this section are conducted on different databases as introduced in Section~\ref{sec:back_dataset}. 
Moreover, the datasets are shuffled to improve the learning process by avoiding any visible or invisible bias induced during the data collection phase. 
This step has been considered optional in the SCA-related literature~\cite{ascad_paper}, although from ML point of view, it is helpful to incorporate into the learning process. 
All the experiments are run, without any special pre-processing of the datasets, on a high-computing cluster with a total of $8$~CPUs allocated per task and a total memory of $80$~GB. The CPUs are the Skylake Dell~$C6420$ model with Xeon Gold~$6142$ processors. 
The rank function provided in the ASCAD package~\cite{ascad_paper} is used in all the experiments described in this section. 


\vspace{0.25ex}
\noindent\textbf{Training and testing sub-dataset Preparation:} 
First, to address the imbalance in the dataset, we create new balanced datasets by performing a data-level method, namely the random undersampling~\cite{related_works:SCA_metrics}. 
This means that only the maximum possible number of data equally for each class is used. 
For instance, for ASCAD\SP{fixed} dataset, we found this number to be $150$, i.e., $150$ traces per class are used to create new balanced datasets. 
Interestingly, InfoNEAT is able to recover the sub-key even when the number of traces per sub-key is reduced (see Section~\ref{sec:results_sca}). 
\ndss{Therefore, it is not needed to rely on oversampling that may lead to overfitting~\cite{related_works:SCA_metrics}. }
After dealing with the imbalance in the dataset, \majorrevision{for datasets with a fixed key for both training and testing data,} we perform the \textit{$k$-fold cross-validation} (see Section~\ref{sec:methodology_training}); otherwise, the trained model is applied against a separate testing set.  
Regardless of whether $k$-fold cross-validation is applied, the training set is used for training the sub-models as well as the stacked model. 
To train the stacked model, a 9:1 split of the training batch size is used as suggested in the literature~\cite{susan2021evaluating}. 

\begin{table}[t]
\centering
\caption{\majorrevision{InfoNEAT parameters to be set by the user.}}
\label{tab:NEAT_parameters}
\resizebox{0.9\columnwidth}{!}{%
\begin{tabular}{|c|c|c|} 
\hline
\textbf{Parameters} & \textbf{Description} & \textbf{Values} \\ 
\hline
\multicolumn{3}{|c|}{\textbf{Parameters related to evaluation of multiple distinct networks}} \\ 
\hline
Population size~($N$) & Number of genomes or networks considered in a generation. & 16 \\ 
\hline
\begin{tabular}[c]{@{}c@{}}Initial Compatibility \\ threshold~($t_c$)\end{tabular} & \begin{tabular}[c]{@{}c@{}}Individuals whose genomic distance is less than this \\ threshold are considered to be part of the same species.\end{tabular} & 1.8 \\
\hline
\multicolumn{3}{|c|}{\textbf{Initialization of weights and biases - using Xavier technique}} \\ 
\hline
\begin{tabular}[c]{@{}c@{}}Weights and biases\\ \{init\_mean,  init\_variation\\min, max\}\end{tabular} & \begin{tabular}[c]{@{}c@{}}\textit{Init\_mean} and \textit{init\_variation} refer to the  Xavier-based distribution\\ of the weights at the start of the algorithm.\\Weights and biases beyond \{min, max\} range are clamped. \end{tabular} & \begin{tabular}[c]{@{}c@{}}\{0, \textit{init\_variation},\\-3$\times$\textit{init\_variation}, \\ 3$\times$\textit{init\_variation}\}\end{tabular} \\ 
\hline
\end{tabular}
}
\end{table}

\subsubsection{Network and Configuration Parameters}
\label{sec:results_neat_architecture}
As discussed in Section~\ref{sec:neat}, the resultant NNs evolved using NEAT are usually quite small (in terms of the number of layers and nodes), yet still quite effective depending on the problem at hand. 
\majorrevision{To obtain such NNs, some parameters as summarized in Table~\ref{tab:NEAT_parameters} should be set before training the NNs through InfoNEAT. 
After setting these, no change in them is required during the evolution of NNs. 
In other words, per learning task (i.e., per dataset), these parameters are set once and remain unchanged (see Appendix~B for more details). 
\ndss{Furthermore, for output nodes, we have selected the \textit{softmax} as our activation function which is one of the commonly used output layer activation function in multi-class classification problems. Softmax assigns probability values to each class or output node such that the sum of these probabilities is $1$.
We apply the numerically stable softmax proposed in~\cite{goodfellow2016deep} in order to deal with the overflow issue (i.e., exploding gradient problem) as observed in ML- and SCA-related studies~\cite{goodfellow2016deep,kerkhof2021no}.}

\vspace{0.25ex}
\noindent\textbf{Parameters related to the initialization of weights and biases, and the batch size:}  These should be chosen carefully by the user, as explained in Appendix B (see  Table~\ref{tab:NEAT_parameters}). 
For ASCAD\SP{fixed} dataset, for instance, the batch size is 150. 
Note that this should be done for \emph{any} deep learning task regardless of the algorithm to be employed. 
After the initialization, the weights and biases are tuned by InfoNEAT over generations. 
Additionally, what is promised and offered by InfoNEAT is the automatic selection of the configuration and tuning of the hyperparameters, including the number of hidden layers, the number of nodes per layer, etc. 
Comparing these to the parameters to be tuned by the user (only the ones in Table~\ref{tab:NEAT_parameters}), we can observe a drastic reduction in user effort. }

\vspace{0.25ex}
\noindent\textbf{Parameters related to evaluation of multiple distinct networks: }
\textit{Population size} can be set based on how many NNs the user wants to evaluate at the same time. Since most of the experiments were conducted on a $16$-core machine, we set this to $16$ to efficiently evaluate multiple NNs.
\ndss{The higher the population size is, the longer a generation takes. 
Intuitively, this value shows how many NNs are evaluated simultaneously.}
\textit{Initial compatibility threshold} is a NEAT parameter that helps maintain diversity in a population of genomes or NNs by comparing the genomic distance of the network with this value (see Section~\ref{sec:neat}). 
\ndss{Here diversity means how different NNs are configured and tuned in a generation. 
Larger thresholds result in having less number of species, i.e., if the population size is small, a modest threshold should be chosen (see~\cite{omelianenko2019hands} for more information). 
Note that the user only defines the initial value of this threshold, and NEAT adjusts that in the next generations automatically.}


\begin{figure}[t]
\centering
\includegraphics[width=0.77\columnwidth]{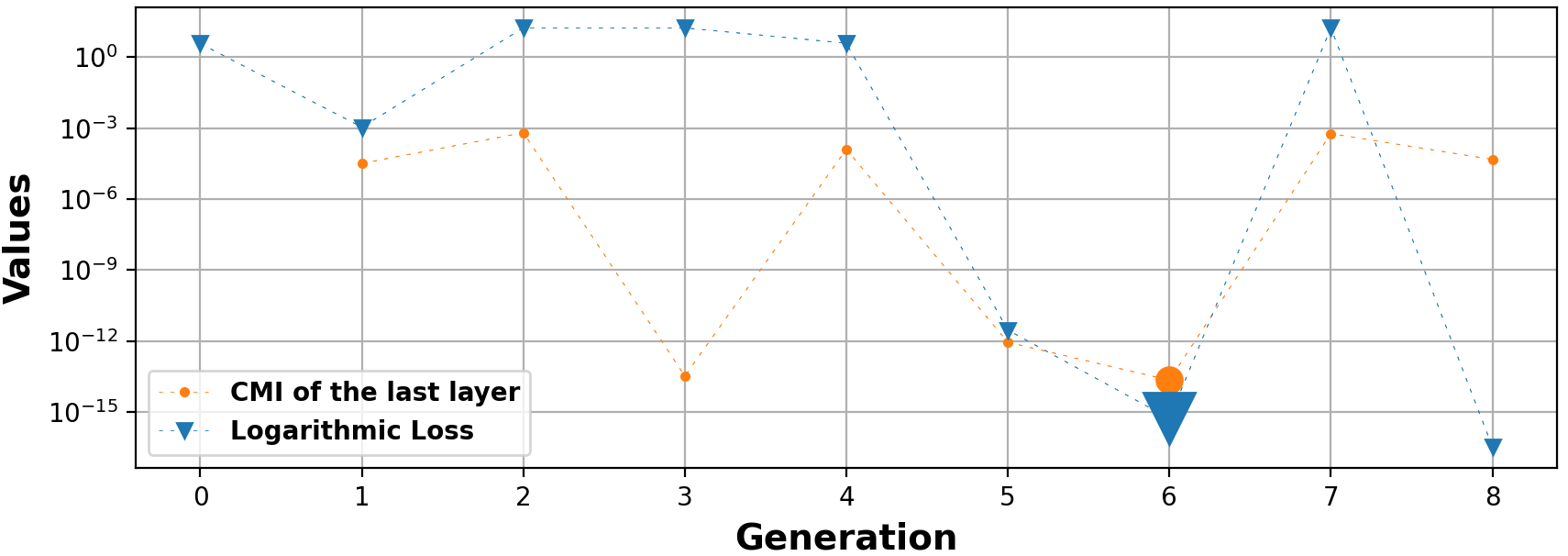}
\caption{Log loss and CMI-based values for the last layer versus different generation. These values are obtained for the best genome at each generation. For better readability, the results for one randomly selected class from ASCAD\SPSB{fixed}{sync} dataset is presented. 
\ndss{Note that if the CMI-based criterion has not been considered, the learning might have been stopped in generation~3, where the fitness (log-loss) value does not decrease; however, the reduction in the CMI indicates that the model is still learning. }
\majorrevision{Dashed lines are drawn for the sake of readability. }}
\label{fig:results_CMI_generation}
\end{figure}

\begin{figure}[t]
\centering
\includegraphics[width=0.8\columnwidth]{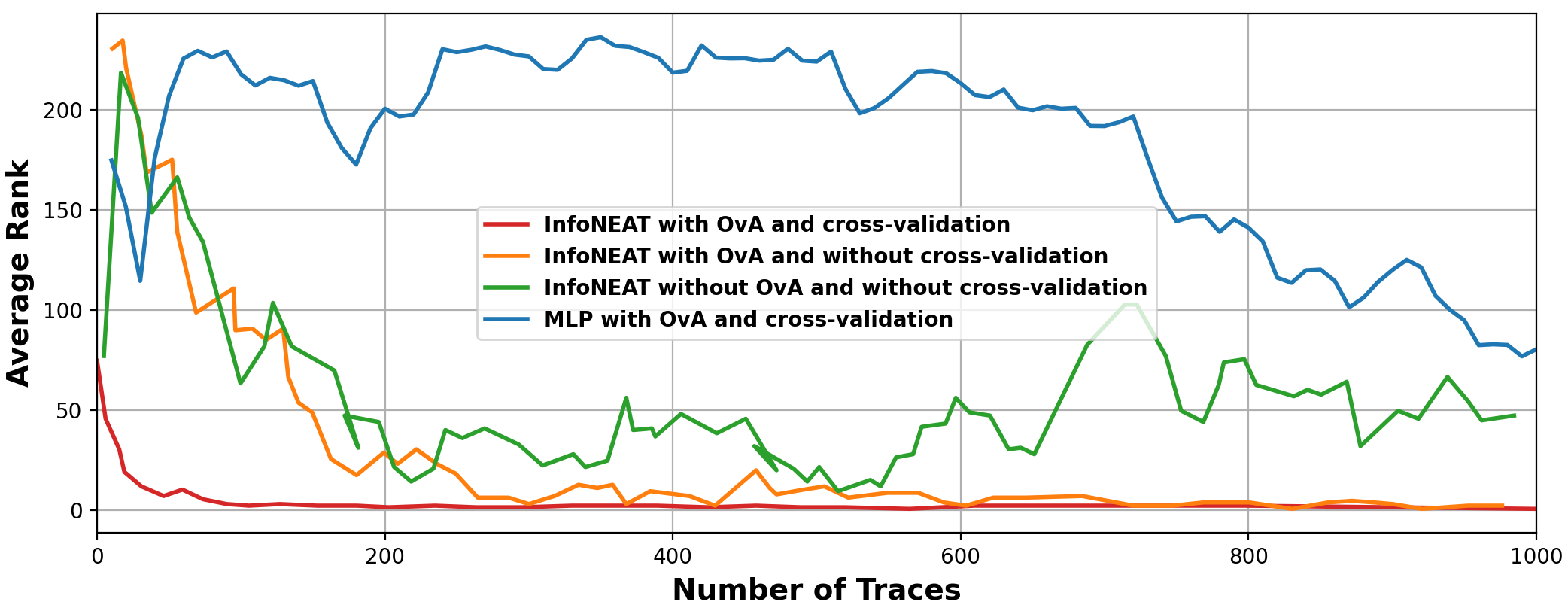} 
\caption{Comparison of average rank obtained for different models obtained with and without stacking and cross-validation. For this experiment, ASCAD\SPSB{fixed}{sync} dataset is used. }
\label{fig:results_stacking}
\end{figure}


\subsection{Examining the Characteristics of InfoNEAT}

\noindent\textbf{Impact of CMI-based Criteria: }
\newversion{As explained in Algorithm~\ref{alg:selection_criteria}, to select the best genome, using log loss as a metric would not be sufficient. 
This is due to the subtle difference between the configurations of NNs in a species that can result in obtaining more than one NN with the same performance. }
Figure~\ref{fig:results_CMI_generation} shows the progression of both the fitness and CMI values for the best genome (or network) during each generation for a randomly chosen class. 
We also employ the CMI-based criteria to stop the evolution process at the right time as described in Algorithm~\ref{alg:stopping_criteria}. 
Whenever the CMI value and log loss both start to degrade, then we select the best genome from the previous generation as our ultimate model for the class. Thus, in Figure~\ref{fig:results_CMI_generation}, we select our ultimate model from generation $6$ rather than generations $7$ or $8$.

\vspace{0.25ex}\noindent\textbf{Training an Effective Stacked Model:}
After the configuration parameters are set and the weights and biases appropriately initialized, we now enter the \textit{training phase}. As discussed in Section~\ref{sec:results_experimental_setup}, we train $256$ different sub-models for the $256$ classes using the balanced ASCAD databases and by employing the OvA technique. 
For the experiment on ASCAD\SP{fixed}, the stacked model is trained using 20 traces per class as discussed before. 
As shown in Figure~\ref{fig:results_stacking}, with stacking and cross-validation involved, the test metric (the average rank in our experiments) improves drastically compared to without stacking and cross-validation.

\begin{table}[t]
\small
\centering
\caption{Comparison of the time taken for InfoNEAT versus NEAT to train a sub-model for each ASCAD\SP{fixed} dataset.}
\label{tab:time_complexity}
\resizebox{0.75\columnwidth}{!}{%
\begin{tabular}{|c|c|c|c|c|} 
\cline{1-5}
& \multicolumn{2}{c|}{\textbf{With InfoNEAT}} & \multicolumn{2}{c|}{\textbf{With NEAT}} \\ 
\hline
\begin{tabular}[c]{@{}c@{}}\textbf{ASCAD\SP{fixed}}\\\textbf{ dataset }\end{tabular} & \begin{tabular}[c]{@{}c@{}}\textbf{Average}\\\textbf{ no. of}\\\textbf{ generations }\end{tabular} & \begin{tabular}[c]{@{}c@{}}\textbf{Average }\\\textbf{ generation}\\\textbf{ time [s] }\end{tabular} & \begin{tabular}[c]{@{}c@{}}\textbf{Average}\\\textbf{no. of}\\\textbf{generations}\end{tabular} & \begin{tabular}[c]{@{}c@{}}\textbf{Average}\\\textbf{generation}\\\textbf{time [s]}\end{tabular} \\ 
\hline
Sync & 8 & 67.384 & 17 & 91.426 \\ 
\hline
Desync50 & 9 & 69.805 & 19 & 103.754 \\ 
\hline
Desync100 & 11 & 68.361 & 20 & 120.682 \\
\hline
\end{tabular}
}
\end{table}

\majorrevision{\vspace{0.25ex}\noindent\textbf{Impact of Irregular Architecture of NNs: } 
To give a better understanding of different aspects of InfoNEAT, especially the importance of architecture delivered by our algorithm, we have designed the following experiment. 
Similar to InfoNEAT, the OvA strategy is used to make 256 binary classification tasks from the multi-class (i.e., 256-class) classification problem underlying the SCA. 
Each binary task associated with a sub-key is handled by training an MLP on a set of traces from the corresponding class and other classes as performed in experiments on InfoNEAT. 
In order to provide a comparison, the number of hidden layers and nodes in the MLPs are set equal to the maximum of those in the NNs generated by InfoNEAT. 
Moreover, the number of epochs, number of traces in a batch, and number of training traces are similar to what has been set for InfoNEAT. 
The average rank presented in Figure~\ref{fig:results_stacking} shows a clear difference between InfoNEAT and MLPs combined through OvA strategy. 
This result indeed indicates the major role played by the irregular configuration of the NNs generated by InfoNEAT.

}
\vspace{0.25ex}\noindent\textbf{Time-complexity for InfoNEAT Training: }
All the steps mentioned in Section~\ref{sec:results_experimental_setup}, such as the configuration of NEAT parameters, initialization of weights and batch size, and the inclusion of CMI-based criteria are part of the InfoNEAT algorithm, and they highly contribute to the improvement of the time-complexity associated with training an effective SCA model for different ASCAD datasets. Table~\ref{tab:time_complexity} summarizes the average time involved in training a particular sub-model using the InfoNEAT algorithm (column labeled as ``With InfoNEAT'') and compares this time to the average time taken while using the default NEAT without the CMI-based criteria (column labeled as ``With NEAT''). 
\majorrevision{For the default NEAT equipped with OvA, we select the batch size of $200$ as recommended in~\cite{ascad_paper}, whereas it is $150$ for InfoNEAT (the reason is explained in Appendix~B). 
InfoNEAT could outperform in terms of the number of generations thanks to the CMI-based best genome selection. 
The matrix-based operations for CMI computation are fast, and the difference in the average generation time is mainly due to the difference in the batch size} (for more discussion and results, see Appendix~C). 


\subsection{Side-channel Analysis through InfoNEAT}
\label{sec:results_sca_metrics}
\subsubsection{ASCAD\SP{fixed} Dataset}\label{sec:results_sca}

\begin{figure}[t]
\centering
\includegraphics[width=0.8\columnwidth]{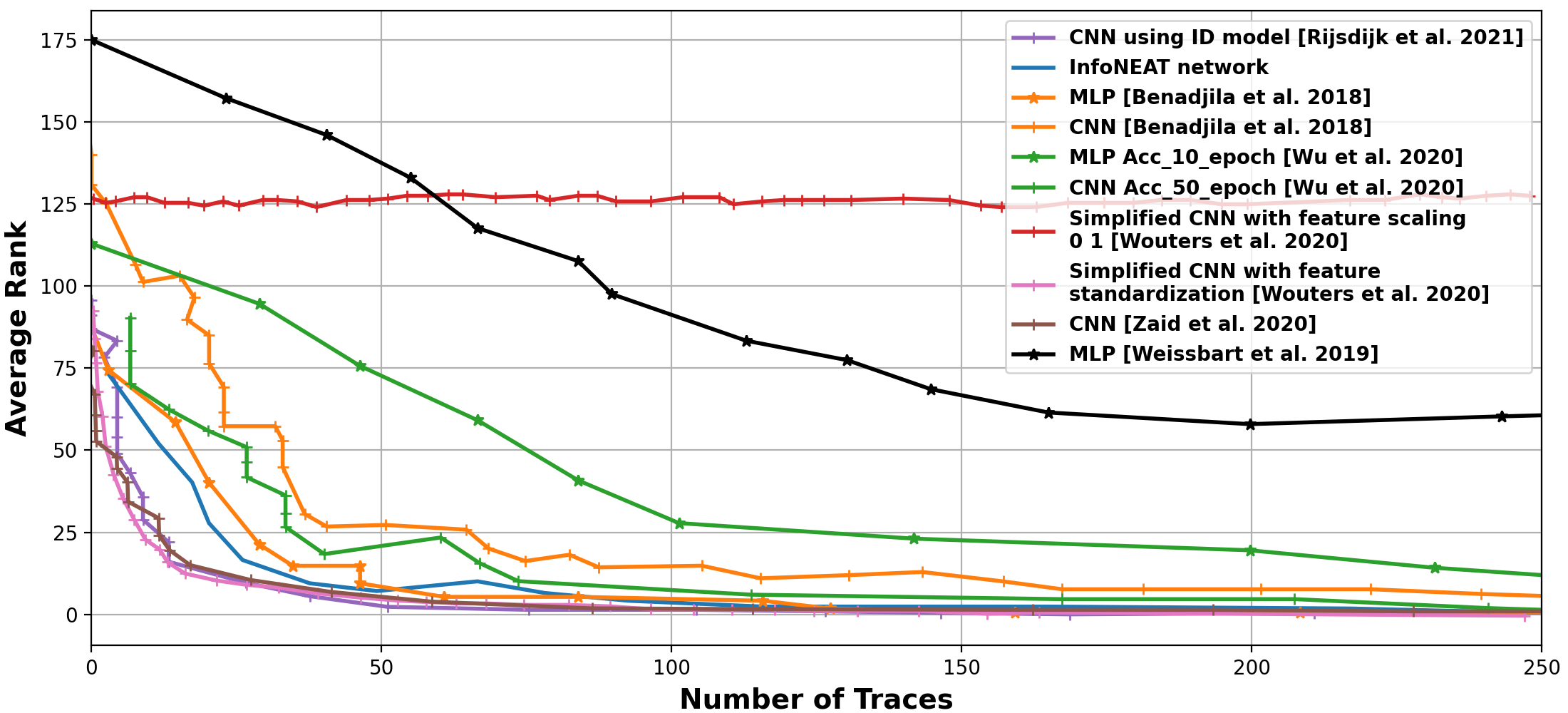} 
\caption{SOTA NNs vs. InfoNEAT applied against ASCAD\SPSB{fixed}{sync}}
\label{fig:results_neat_comparison}
\end{figure}

Figure~\ref{fig:results_neat_comparison} shows the comparison of SOTA networks and InfoNEAT for the ASCAD\SPSB{fixed}{sync} dataset. 
\ndss{For the results presented in~\cite{related_works:automated_hyperparameter_tuning}, the best results for MLPs and CNNs are depicted. 
Since feature scaling between 0 and 1 is applied in InfoNEAT, in addition to the best result from~\cite{Wouters_Arribas_Gierlichs_Preneel_2020}, the result for the similar scaling is also presented. 
It is remarkable that compared with the most relevant approaches, applying MLPs with hyperparameter tuning~\cite{related_works:automated_hyperparameter_tuning}, InfoNEAT's attack performance is slightly better. 
Considering CNNs with hyperparameter tuning~\cite{zaid2020methodology,related_works:hyperparameter_tuning_RL,Wouters_Arribas_Gierlichs_Preneel_2020}, InfoNEAT outperforms the 
model in~\cite{Wouters_Arribas_Gierlichs_Preneel_2020} trained on traces, whose features are scaled between 0 and 1 (similar to InfoNEAT). 
For this type of feature-scaling, the model in~\cite{zaid2020methodology} and InfoNEAT yield similar performance. 
InfoNEAT and the model trained on standardized features (zero-mean unit-variance) in~\cite{Wouters_Arribas_Gierlichs_Preneel_2020} show similar performance. 
\minr{In a nutshell, comparing InfoNEAT's results with the best results in~\cite{zaid2020methodology,Wouters_Arribas_Gierlichs_Preneel_2020}, they have presented  $T_{GE0}=191, 155$, respectively, whereas InfoNEAT achieves $T_{GE0}=130$, where $T_{GE0}$ denotes the least number of attack traces required to break the target. } 
Nevertheless, InfoNEAT's largest potential for improving SCA performance becomes evident when ASCAD\SB{desync} dataset is considered. }  

\begin{figure}[t]
\centering
\includegraphics[width=0.8\columnwidth]{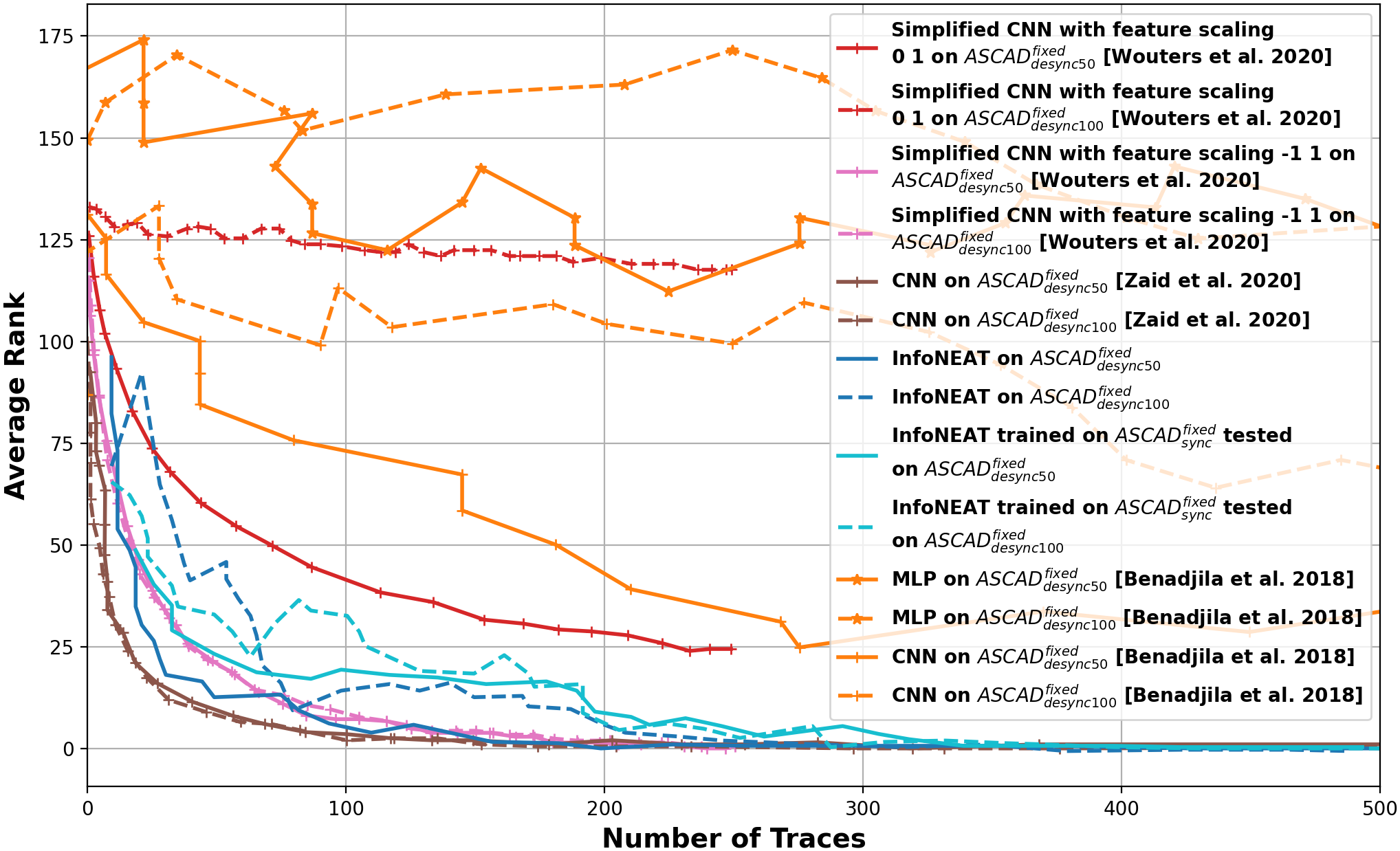} 
\caption{Average rank for the ASCAD\SPSB{fixed}{desync} dataset.}
\label{fig:results_neat_all_dataset}
\end{figure}

\vspace{5pt}\ndss{\noindent\textbf{Training on ASCAD\SPSB{fixed}{sync} and testing against ASCAD\SPSB{fixed}{desync}:}} 
Figure~\ref{fig:results_neat_all_dataset} show the results of using InfoNEAT model against different ASCAD\SPSB{fixed}{desync} dataset. 
As can be seen, InfoNEAT is capable of learning different desynchronized datasets when trained and tested on the same dataset.  
\minr{As it might not always be easy to align traces using alignment techniques (see,e.g.,~\cite{won2021back}), it is useful to employ ML models robust against desynchronization~\cite{related_works:plaintext,related_works:MI_SCA}. 
In our case, once the model is trained against synchronized traces (during the profiling phase where the attacker works on an open copy), there is no need to re-train/fine-tune the model, even in the presence of desynchronization.}
More interestingly, the average rank computed by using the model trained on synchronized data to attack or test the desynchronized data is comparable to the other average ranks (trained and tested on the same type of dataset). 
In this case, this shows that the models trained using InfoNEAT are \textit{generalizable} in the sense that they are not sensitive to desynchronization/jittering (see also the results for ASCAD\SPSB{var}{desync} in Section~\ref{sec:results_sca_ascad_var}). 
\majorrevision{This is due to the non-typical NN architecture created by NEAT that can handle time-dependent, or in general, order-dependent phenomena~\cite{caamano2015tau}.  
This is made possible thanks to the feedback connections between neurons automatically inserted using the add connection operator. In other words, networks generated and trained by InfoNEAT can adapt to sequences of events, even if they exhibit a lag, e.g., time-delay. }

\begin{table*}[t]
\small
\centering
\caption{\ndss{Comparison between InfoNEAT and SOTA attacks against ASCAD\SP{var} dataset. \vspace{2pt}}}
\label{tab:varkey_results}
\resizebox{\textwidth}{!}{%
\begin{tabular}{|c|c|c|c|c|c||c|c|} 
\hline
\multirow{1}{*}{\textbf{Metric}} 
& \multirow{1}{*}{\textbf{\cite{wu2021best}}} &
\multirow{1}{*}{\textbf{\cite{related_works:plaintext}}}
& \multirow{1}{*}{\textbf{\cite{related_works:ensemble_SCA}}}
& \multirow{1}{*}{\textbf{\cite{related_works:automated_hyperparameter_tuning}}}
& \multirow{1}{*}{\textbf{\cite{related_works:hyperparameter_tuning_RL}}}
& \multirow{1}{*}{\textbf{\cite{related_works:least_num_traces}}}
&\multicolumn{1}{c|}{\textbf{InfoNEAT}}\\

\hline
$T_{GE0}$ & 188$\dag$ & F & 56*~/~450$\dag$ & 2000*~/~3144$\dag$ & 490$\dag$ & 1040*~/~2854$\dag$ & \textbf{120}\\
\hline
$T_{GE1}$ & 81$\dag$ & F$\dag$ & 50*~/~400$\dag$ & 1250*~/~3000$\dag$ & 265$\dag$ & NR & \textbf{30}\\
\hline
$T_{GE20}$  & 40$\dag$ & 6$\dag$ & 6*~/~50$\dag$ & 150*~/~100$\dag$ & 38$\dag$ & 56*~/~318$\dag$ & \textbf{10}\\
\hline
\end{tabular}
}
\begin{tablenotes}
\item[*] \footnotesize{ $*$ MLP, $\dag$ CNN, \hspace{10pt} F: failed to reach \hspace{10pt} NR: Not reported.\vspace{10pt}} 
\end{tablenotes}
\end{table*}

\begin{table*}[t]
\small
\centering
\caption{\ndss{Qualitative comparison between important characteristics of the methodologies applied against ASCAD\SP{var} dataset. } \vspace{2pt}}
\label{tab:varkey_quality}
\resizebox{\textwidth}{!}{%
\begin{tabular}{|c|c|c|c|c|c||c|c|} 
\hline
\multirow{1}{*}{\textbf{Required/used}} 
& \multirow{1}{*}{\textbf{\cite{wu2021best}}} &
\multirow{1}{*}{\textbf{\cite{related_works:plaintext}}}
& \multirow{1}{*}{\textbf{\cite{related_works:ensemble_SCA}}}
& \multirow{1}{*}{\textbf{\cite{related_works:automated_hyperparameter_tuning}}}
& \multirow{1}{*}{\textbf{\cite{related_works:hyperparameter_tuning_RL}}}
& \multirow{1}{*}{\textbf{\cite{related_works:least_num_traces}}}
&\multicolumn{1}{c|}{\textbf{InfoNEAT}}\\

\hline
Plaintext & \textcolor{mygreen}{No} & \textcolor{red}{Yes} &\textcolor{mygreen}{No} &  \textcolor{mygreen}{No}& \textcolor{mygreen}{No} & \textcolor{mygreen}{No} & \textcolor{mygreen}{No}\\
\hline
Template attack & \textcolor{red}{Yes} & \textcolor{mygreen}{No} & \textcolor{mygreen}{No} & \textcolor{mygreen}{No} & \textcolor{mygreen}{No} & \textcolor{mygreen}{No}& \textcolor{mygreen}{No} \\
\hline
Additional model validation step  & \textcolor{mygreen}{No} & \textcolor{red}{Yes} & \textcolor{red}{Yes} & \textcolor{red}{Yes}  & \textcolor{red}{Yes} & \textcolor{mygreen}{No}& \textcolor{mygreen}{No}\\
\hline
\end{tabular}
}
\end{table*}

\begin{table}[t]
\small
\centering
\caption{\ndss{NNs trained on ASCAD\SPSB{var}{sync} against ASCAD\SPSB{var}{desync}}. \vspace{2pt}}
\label{tab:varkey_results_desync}
\resizebox{0.75\columnwidth}{!}{%
\begin{tabular}{|c|c c||c c|c c|} 
\hline
\multirow{2}{*}{\textbf{Metric}} 
&\multicolumn{2}{c||}{\multirow{1}{*}{CNN~\cite{related_works:plaintext}}} 
&\multicolumn{2}{c|}{\multirow{1}{*}{VGG16~\cite{ascad_paper}}}
&\multicolumn{2}{c|}{\textbf{InfoNEAT}}\\
\cline{2-7}
&\begin{tabular}[c]{@{}c@{}}desync50\end{tabular}
&\begin{tabular}[c]{@{}c@{}}desync100\end{tabular}
&\begin{tabular}[c]{@{}c@{}}desync50\end{tabular}
&\begin{tabular}[c]{@{}c@{}}desync100\end{tabular}
&\begin{tabular}[c]{@{}c@{}}desync50\end{tabular}
&\begin{tabular}[c]{@{}c@{}}desync100\end{tabular}\\
\hline

$T_{GE0}$ &F & F  & F & F & 880 & 960 \\
\hline
$T_{GE1}$ & F & F &  F & F & 590 & 140 \\
\hline
$T_{GE20}$ & F & F & 300 & F & 170 & 70\\
\hline
$T_{GE50}$ & 935 & F & 90 & F & 70 & 30 \\
\hline
\end{tabular}
} 
\begin{tablenotes}
\item[*] \footnotesize{ desync50 and desync100 denote ASCAD\SPSB{var}{desync50} and ASCAD\SPSB{var}{desync100}, respectively. F: failed to reach}
\end{tablenotes}
\end{table}




\ndss{
InfoNEAT's performance in the face of more challenging datasets is considered next, where it showcases its potential even better.}


\majorrevision{
\subsubsection{ASCAD\SP{var}Dataset}\label{sec:results_sca_ascad_var}
In the same vein as ASCAD\SP{fixed}, the batch size is set as explained in Appendix~B. 
For ASCAD\SPSB{var}{sync} dataset, it is determined that the batch size of 256 is an adequate choice. 
\ndss{We performed the experiments with 20 and 25 traces per class and observed no significant improvement if the latter is chosen; hence, the results for the former are reported here. }

\ndss{For SCA, in the case of fixed key, $k$-fold cross-validation has been employed as a powerful tool for making a more realistic estimate of the model's skill than other methods, such as a simple train/test split~\cite{ascad_paper,related_works:SCA_metrics,kim2019make,bhasin2020mind}. 
For variable key cases, to report a good estimation of the average rank, the attack should be launched multiple times to assess the attack performance properly~\cite{wu2020attack,wu2022evaluation}. 
This is due to the fact that if a number of independent experiments (key rank evaluations) are conducted in the attack phase, the attack performance can vary. 
Although this variation might be small, it is needed to mount the attack multiple times (e.g., around 40 independent attacks~\cite{wu2022evaluation}) using a model with properly optimized hyperparameters, as for InfoNEAT.  
To account for the performance variation, summary statistics should be taken into account, e.g., the median has been suggested to be the most effective in this respect~\cite{wu2022evaluation}; hence, our results present the median, in addition to the minimum, calculated over 50 independent attacks. 

In Table~\ref{tab:varkey_results}, the least number of attack traces required to break the target ($T_{GE0}$), obtain an average rank $1$ ($T_{GE1}$), and $<20$ ($T_{GE20}$) are provided. 
Note that the results of other studies are directly derived from their papers, and the \emph{minimum} number of traces for different metrics is provided. 
Noteworthy is that ~\cite{related_works:least_num_traces} has not reported $T_{GE1}$. 
\minr{In this table, ``F'' indicates that according to the results in the respective papers, the proposed attack could not reach the average rank of 0, 1, or below 20. 
We should add that these results have been obtained for the specific setting considered in~\cite{related_works:plaintext}. 
Their model is trained using ASCAD\SP{var} training dataset, whereas it is tested in 100 runs, each with 100 traces, resulting in a total of 10,000 testing traces. 
The results in~\cite{related_works:plaintext} are the average over these 100 runs. 
We stress that although their attack could not reach the average rank 0 or 1, it achieved rank 2 with only 40 traces, comparable to InfoNEAT, which reaches the average rank 1 with 30 traces. }

\cite{related_works:ensemble_SCA} reported one of the best results not only for the ensemble of models, but also for the \emph{best} model. 
In their framework, the best model(s) is selected with regard to the validation key rank, i.e., an additional step is taken to fine-tune the model after some epochs by applying the trained model against a validation set which is created using the attack traces. 
This directly affects the performance of the model cf.~\cite{related_works:ensemble_SCA}. 
The same holds for approaches in~\cite{related_works:automated_hyperparameter_tuning, related_works:hyperparameter_tuning_RL,related_works:plaintext}. 
As this additional step enhances the performance of the attack, it is important to differentiate between these approaches and InfoNEAT and~\cite{related_works:least_num_traces}, which do \emph{not} take advantage of the validation step. 
Moreover,~\cite{wu2021best} has combined DL and template attacks, whereas~\cite{related_works:plaintext} has incorporated plaintext as a feature (see Table~\ref{tab:varkey_quality} for a summary). 
Additionally, for InfoNEAT, the \emph{median} values are $T_{GE0}=310$, $T_{GE1}=240$, and $T_{GE20}=145$. 


\vspace{5pt}\noindent\textbf{Training on ASCAD\SPSB{var}{sync} and testing against ASCAD\SPSB{var}{desync}: }
As demonstrated for ASCAD\SP{fixed} dataset, it is possible to train the model on synched traces and apply it against desynchronized traces, thanks to the irregular configuration of NNs evolved by InfoNEAT. 
Interestingly enough, the same observation is made when the model trained on ASCAD\SPSB{var}{sync} is applied against ASCAD\SPSB{var}{desync50} and ASCAD\SPSB{var}{desync100}. 
As the models in~\cite{related_works:plaintext,ascad_paper} could not deliver results aligned with the metrics $T_{GE0}$--$T_{GE20}$, we add $T_{GE50}$ to this table, also considered as a performance metric in the relevant literature cf.~\cite{kim2019make}.
\minr{Table~\ref{tab:varkey_results_desync} compares InfoNEAT's performance with the results presented in~\cite{related_works:plaintext}. 
Similar to Table~\ref{tab:varkey_results}, ``F'' indicates that the attack could not achieve rank 0,1,20, or 50. 
In order to test the model, 10 runs, each with 1000 traces, are performed in~\cite{related_works:plaintext} and the results are the average over the runs. }
\minr{Interestingly, as can be seen in Table~\ref{tab:varkey_results_desync}, InfoNEAT can reach rank 0 within 1000 testing traces, while the minimum rank is 48 obtained for 935 testing traces~\cite{related_works:plaintext} and 10 for 700 traces (using VGG16~\cite{ascad_paper}). }
For InfoNEAT, for ASCAD\SPSB{fixed}{desync50} (ASCAD\SPSB{fixed}{desync100}) the \emph{median} values are $T_{GE0}=970$ ($975$), $T_{GE1}=870$ ($675$), $T_{GE20}=480$ ($395$), $T_{GE50}=310$ ($185$). 
Note that, to the best of our knowledge, no other results for NNs trained on ASCAD\SPSB{var}{sync} to attack ASCAD\SPSB{var}{desync50} and ASCAD\SPSB{var}{desync100} have been reported.} 

\begin{figure}[t]
\centering
\includegraphics[width=0.9\columnwidth]{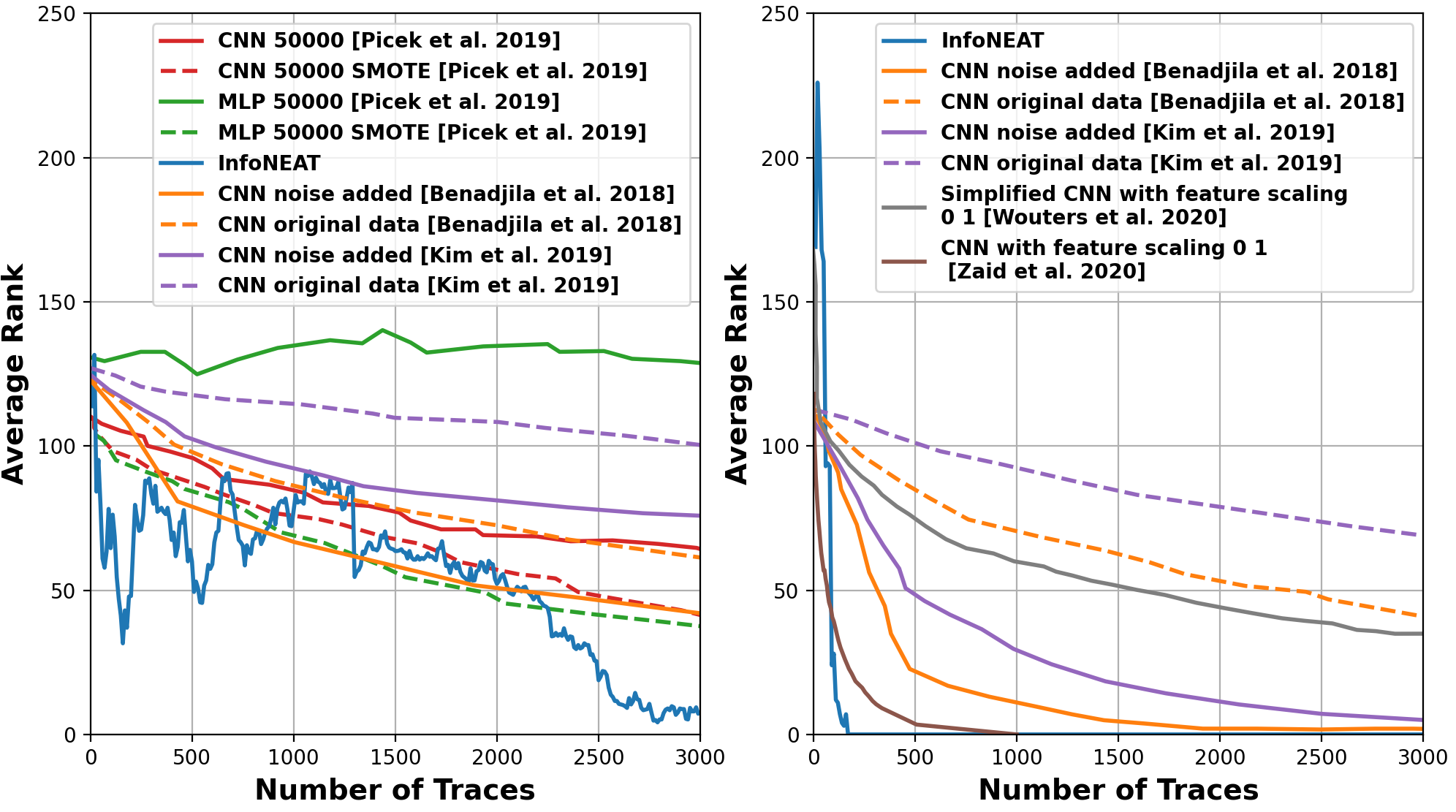} 
\caption{\ndss{SOTA NNs vs. InfoNEAT for AES\textunderscore HD (a) with cross-validation: 
$T_{GE50}$ is 2000 (InfoNEAT), 2000 (MLP with SMOTE and CNN in~\cite{ascad_paper} with noise regularization~\cite{kim2019make}) and 2500 (CNN with SMOTE)~\cite{related_works:SCA_metrics}; 
(b) best models.}}
\label{fig:results_neat_comparison_aeshd}
\end{figure}

\subsubsection{AES\_HD Dataset}\label{sec:results_sca_AES_HD}
To train the NNs, we follow the procedure explained in Appendix~B to determine the size of batches, which is set to 256, whereas the number of traces used to train the stacked model is 30. 
The average number of generations to train a sub-model is 19. 
The number of folds for cross-validation is set to 5, similar to~\cite{related_works:SCA_metrics}. 
\ndss{ 
For the MLP results from this study, no feature scaling has been applied; hence, a direct comparison between our results and ones in~\cite{related_works:SCA_metrics} is not possible. 
For the sake of better comparison, we also take into account their best results obtained after applying SMOTE, i.e., a resampling method that adds synthetic examples with the random shift, as well as applies other transformations. 
Similar to undersampling, SMOTE can deal with imbalanced datasets. 

As shown in Figure~\ref{fig:results_neat_comparison_aeshd}, we compare our results with the SOTA methods based on k-fold cross validation shown in Figure~\ref{fig:results_neat_comparison_aeshd}(a) and the best model (fold) as depicted in Figure~\ref{fig:results_neat_comparison_aeshd}(b). 
This is with regard to the observation made before in the literature~\cite{kim2019make,wu2020attack} and by us during the testing phase.  
For some folds, the average rank decreases and reaches 0 by increasing the number of testing traces, whereas, for others, it decreases and remains below a threshold, e.g., 50 in our experiments. 
Note that the trend observed here and in the prior studies is similar, although the minimum average rank obtained by us is less than what has been reported before. 
Comparing to~\cite{related_works:SCA_metrics} that considers $k$-fold cross validation, $T_{GE50}=2000,2500$ is observed for MLP and CNN, respectively, whereas $T_{GE50}=2000$ in our case (see Figure~\ref{fig:results_neat_comparison_aeshd}(a)). 
It is interesting to compare the trend in the results obtained by using InfoNEAT and the SOTA approaches. 
The latter has shown an almost smooth trend, contrary to InfoNEAT. 
This can be explained by the differences between these methods when being employed against relatively more noisy measurements in the AES\_HD dataset.  
First, algorithms applying SMOTE and adding noise to the traces leverage regulation-like techniques, which are useful in the test phase as well as training~\cite{kim2019make,related_works:SCA_metrics}. 
On top of this, and more importantly, we should stress that the k-fold validation method in~\cite{kim2019make} is different from ours. 
In that study, repeated random subsampling has been used to reduce the instability. Moreover,~\cite{related_works:SCA_metrics} has considered feature extraction on traces. 
InfoNEAT has applied none of these techniques, and they are left for future work. 

Our best model achieves $T_{GE0}=170$, whereas using CNNs combined with pre-processing leads to $T_{GE0}\approx 800$ (averaged over 100 trials)~\cite{Wouters_Arribas_Gierlichs_Preneel_2020,zaid2020methodology} (see Figure~\ref{fig:results_neat_comparison_aeshd}(b)). 
Similarly, multi-scale CNN along with various data transformation is used to obtain $T_{GE50}=2000$ and $T_{GE10}=5000$~\cite{won2021back}, whereas $T_{GE10}=130$ for InfoNEAT. 
Besides those results depicted in this figure, one can consider the ones presented in~\cite{van2020learning}. 
In that work, it has been observed that if a compact network (MLP or CNN), called the student network, is trained on the output of a big (teacher) network, better results can be obtained when compared to complex networks. 
Nevertheless, the minimum number of traces for which the average rank goes below 100 (best case presented) is 10,000.}}

\ndss{\vspace{5pt}\noindent\textbf{Summary of the results obtained for different datasets: }The most important message conveyed here is that InfoNEAT could achieve high attack performance, even \emph{without} combining DL with a template attack or leveraging the plaintext as a feature or model validation. 
The models trained using InfoNEAT also enjoy another advantage, namely being resilient against desynchronization. }


\subsection{Memory and Epoch-wise Efficiency}
\majorrevision{
Table~\ref{tab:results_comparison} compares SOTA networks with InfoNEAT when attacking various datasets. }
\ndss{Note that this table presents the number of \emph{trainable} parameters, different from the number of hyperparameters (for the computational complexity of hyperparameter tuning and NAS, see Section~\ref{sec:results_discussion}).  }
It is worth mentioning that for the sake of comparison, we consider the size of the individual models because we train them one by one. 
\ndss{Similar to the bagging ensemble technique used in~\cite{related_works:ensemble_SCA}, the complexity of the stacked model is linear in the number of models} (see Appendix~C for additional information on the number of nodes and number of species).  
In general, InfoNEAT requires fewer epochs to train an NN. 
\ndss{As can be seen in Table~\ref{tab:results_comparison}, the number of trainable parameters for CNNs is less than that for MLPs and InfoNEAT. 
Compared to MLPs, InfoNEAT shows improvement. 
Note that the comparison with~\cite{related_works:ensemble_SCA} is based on the average number of layers and neurons reported in this study (no concrete number is reported). 
Nevertheless, when comparing the average rank obtained by all types of NNs, they could perform well (see Figures~\ref{fig:results_neat_comparison}-\ref{fig:results_neat_comparison_aeshd} and Tables~\ref{tab:varkey_results}-\ref{tab:varkey_results_desync}). 
Yet, InfoNEAT can achieve the competitive result \emph{without} applying any additional technique \minr{as also summarized in Table~\ref{tab:results_comparison}}. 
It also delivers desynchronization-robust NNs in our experiments. 
Next, we discuss other technical aspects of InfoNEAT and provide a cost-benefit evaluation. }

\begin{table}
\centering
\caption{SOTA approaches vs. InfoNEAT. }
\label{tab:results_comparison}
\resizebox{\linewidth}{!}{%
\begin{tabular}{|c|c|c|c|c|c|c|} 
\hline
\begin{tabular}[c]{@{}c@{}}\\\textbf{Ref.}\end{tabular} & 
\begin{tabular}[c]{@{}c@{}}\textbf{No. of trainable }\\\textbf{ parameters }\end{tabular} & \begin{tabular}[c]{@{}c@{}}\textbf{No. of}\\\textbf{ epochs }\end{tabular} & \begin{tabular}[c]{@{}c@{}}\textbf{No. of}\\\textbf{ training}\\\textbf{ traces }\end{tabular} & \begin{tabular}[c]{@{}c@{}}\textbf{Hyperparameter}\\\textbf{ tuning }\end{tabular} & \textbf{Dataset} & \minr{\textbf{$T_{GE0}$}} \\ 
\hline
\multirow{2}{*}{\begin{tabular}[c]{@{}c@{}}\cite{ascad_paper}\end{tabular}}  & \multirow{2}{*}{\begin{tabular}[c]{@{}c@{}}393,936*\\66,652,444$\dag$ \end{tabular}}
& \multirow{2}{*}{\begin{tabular}[c]{@{}c@{}}400*\\75$\dag$ \end{tabular}}
& \multirow{2}{*}{\begin{tabular}[c]{@{}c@{}}40,000\end{tabular}}
& \multirow{2}{*}{\begin{tabular}[c]{@{}c@{}}Trial and error \end{tabular}} & \multirow{2}{*}{\begin{tabular}[c]{@{}c@{}}ASCAD\SP{fixed} \end{tabular}} & \multirow{2}{*}{\begin{tabular}[c]{@{}c@{}}410*\\480$\dag$ \end{tabular}}\\ 
 &  & &  &  & & \\ 
\hline
\begin{tabular}[c]{@{}c@{}}\cite{related_works:MLP_weissbart}\end{tabular}  & 740,136* & 200 & 20,000 & Trial and error & ASCAD\SP{fixed} & 1630\\ 
\hline
\multirow{2}{*}{\begin{tabular}[c]{@{}c@{}}\cite{related_works:automated_hyperparameter_tuning}\end{tabular}} &\multirow{2}{*}{\begin{tabular}[c]{@{}c@{}} 478,656* \\ 54,752$\dag$\end{tabular}}
& \multirow{2}{*}{{[}10,50]~} & \multirow{2}{*}{50,000} & \multirow{2}{*}{\begin{tabular}[c]{@{}c@{}}Bayesian optimization \\ and Random search\end{tabular}} & ASCAD\SP{fixed} & 129*/158$\dag$\\ 
\cline{6-7}
 &  &  &  &  & ASCAD\SP{var} & 2000*~/~3144$\dag$\\ 
\hline
\cite{related_works:SCA_metrics} & 49,024* & - & 50,000 & Randomly selected & AES\_HD & 10,000\\ 
\hline
\cite{related_works:ensemble_SCA} & 493,480* & - & 200,000 & Random Search & ASCAD\SP{var} & 56*~/~450$\dag$\\ 

\hline

\multirow{2}{*}{\begin{tabular}[c]{@{}c@{}}\cite{zaid2020methodology}\end{tabular}}& 16,960$\dag$ & 50 & 50,000 & \multirow{2}{*}{\begin{tabular}[c]{@{}c@{}} Visualization techniques\end{tabular}} & ASCAD\SP{fixed} & 191$\dag$\\ 
\cline{2-3}\cline{4-4}\cline{6-7}
 & 3,282$\dag$ & 20 & 45,000 &  & AES\_HD & 800$\dag$\\ 
\hline

\multirow{2}{*}{\begin{tabular}[c]{@{}c@{}}\cite{Wouters_Arribas_Gierlichs_Preneel_2020}\end{tabular}} & 6,436$\dag$ & \multirow{2}{*}{50} & 45,000 & \multirow{2}{*}{\begin{tabular}[c]{@{}c@{}} Taken from\cite{zaid2020methodology}\end{tabular}} & ASCAD\SP{fixed} & 155$\dag$\\ 
\cline{2-2}\cline{4-4}\cline{6-7}
 & 2,020$\dag$ &  & 50,000 &  & AES\_HD & 800$\dag$\\ 
\hline

\multirow{2}{*}{\begin{tabular}[c]{@{}c@{}}\cite{related_works:hyperparameter_tuning_RL}\end{tabular}} & 79,439$\dag$ & \multirow{2}{*}{50} & 50,000 & \multirow{2}{*}{\begin{tabular}[c]{@{}c@{}}Reinforcement learning\end{tabular}} & ASCAD\SP{fixed} & 202$\dag$\\ 
\cline{2-2}\cline{4-4}\cline{6-7}
 & 70,492$\dag$ &  & 200,000 &  & ASCAD\SP{var} & 490$\dag$\\

\hline
\multirow{3}{*}{InfoNEAT} & 15,107 & 8 & 38,400 & \multirow{3}{*}{\begin{tabular}[c]{@{}c@{}}Automatic \\ Neuroevolution\end{tabular}} & ASCAD\SP{fixed} & 130\\ 
\cline{2-4}\cline{6-7} & 317,408 & 8 & 161,280 &  & ASCAD\SP{var} & 120\\ 
\cline{2-4}\cline{6-7} & 102,757 & 33 & 38,400 &  & AES\_HD & 170\\
\hline

\end{tabular}
}
\begin{tablenotes}
\item[*] \footnotesize{ $*$ MLP, $\dag$ CNN. } 
\end{tablenotes} 
\end{table}
\ndss{
\subsection{Discussion}\label{sec:results_discussion}
\vspace{5pt}\noindent\textbf{InfoNEAT vs. approaches for hyperparameter optimization:}
When it comes to tuning the hyperparameters, the main proposals in SCA-related literature include random search and Bayesian optimization. 
Random search evaluates uniformly random points in the hyperparameter space to select the one that provides the best performance. 
This implementation is easier to comprehend and parallelize. 
Despite its ease of use, random search is burdened by the high variance between runs. 
Moreover, from one run to another, this algorithm cannot use previous observations leading to a long convergence time in some scenarios. 
An example of this is reported in~\cite{wu2021best} when unsuccessfully mounting the ensemble learning attack~\cite{related_works:ensemble_SCA} with a non-optimal hyperparameter space caused by the input dimensions. 
If the number of runs is restricted to avoid this, random search may not deliver the optimum result~\cite{andradottir2015review}. 

Another promising candidate for hyperparameter tuning is Bayesian optimization, which is, on one hand, data-efficient, but on the other hand, computationally expensive. 
Concretely, for a given number of evaluations of candidate solution $N$, its complexity is $O(N^3)$, and even after applying mechanisms devised to reduce this cost, it grows substantially with the number of evaluations~\cite{lan2022time}. 
\minr{The complexity of InfoNEAT is similar to the NEAT algorithm. 
This is due to the fact that while the stopping and best-genome selection criteria can reduce the number of candidates to be evaluated per submodel, the maximum number of evaluations does not exceed that of the NEAT algorithm. 
It is also obvious that when training all submodels, since the evaluation of candidate solutions is repeated for a constant number (i.e., 256 for SCA), the complexity remains the same compared to NEAT. 
Therefore, the complexity of the InfoNEAT algorithm is  $O(N^2)$, similar to NEAT~\cite{lobo2000time,baldominos2020automated}. }
Given this complexity, Bayesian optimization could take longer to converge to the solution. 
Nevertheless, Bayesian optimization is more efficient than Grid search with the complexity $O(N^{k})$, which is used in, e.g.,~\cite{related_works:MLP_weissbart}. 
Moreover, Bayesian optimization follows a more systematic search mechanism compared to random search~\cite{lorenzo2017particle}. 

\vspace{5pt}\noindent\textbf{InfoNEAT vs. approaches for NAS (beyond hyperparameter optimization): }
When formalizing NAS as a reinforcement learning task, different RL approaches with various policies and optimization techniques could be taken into account~\cite{elsken2019neural}. 
Baker et al.'s Q-learning algorithm~\cite{baker2016designing} is one of such proposals that has found application in SCA as well~\cite{related_works:hyperparameter_tuning_RL}. 
\cite{baker2016designing} has proposed to train a policy that sequentially chooses not only the type of each layer, but also its corresponding hyperparameters. 
The main drawback of such a technique is its high exploration cost and slow convergence time~\cite{baker2016designing,wang2021robust,related_works:hyperparameter_tuning_RL}. 
Precisely, the complexity of Q-learning proposed in~\cite{baker2016designing} is $\tilde{\mathcal{O}}(T)$, where $T$ is the total number of steps and $\tilde{\mathcal{O}}(\cdot)$ denotes that $T \geq \text{polylog}(S, A, H) $ with $S$, $A$, and $H$ denoting the number of states, actions, and steps. 
This means that the complexity depends on the number of states, actions, and steps up to a poly-log degree. 
Although a direct comparison with InfoNEAT computational complexity may not be possible, InfoNEAT could exhibit less computational complexity in practical scenarios as demonstrated for NEAT cf.~\cite{galvan2021neuroevolution}. 

\vspace{5pt}\noindent\textbf{Scalability of InfoNEAT:} 
Neuroevolutionary approaches are alternative solutions to RL, where recent case studies have demonstrated their comparably-well performances~\cite{real2019regularized}. 
In addition, neuroevolution is more advantageous because it avoids backpropagation and has comparatively less time complexity, although one challenge that remains to be addressed is the scalability. 
Neuroevolution at scale is an active research area, with many methods being inspired by NEAT's ability to increase complexity over generations. 
In this regard, adding layers instead of individual neurons is one of the promising solutions~\cite{rawal2018nodes,real2019regularized}. 
Our observation is that InfoNEAT taking advantage of meta-learning, can generate compact networks that are effective under SCA scenarios, in line with the conclusion made in~\cite{related_works:automated_hyperparameter_tuning,related_works:hyperparameter_tuning_RL}. 
Our future work will also consider other variants of neuroevolution. 

\vspace{5pt}\noindent\textbf{Is InfoNEAT a silver bullet?}
InfoNEAT should be considered another step toward fostering research into the application of NAS in SCA. 
In fact, methods devised in~\cite{related_works:hyperparameter_tuning_RL} and our paper are computationally expensive, although the effort made by the attacker/evaluator launching SCA and the reliance on their own expertise can be reduced. 
From a more technical perspective, NAS and especially InfoNEAT bring the advantage of configuring NNs that human experts might not easily develop manually. 
From this perspective, InfoNEAT could demonstrate its potential, in particular, when more challenging datasets have been considered. 
Our results show that such networks and their applications in SCA can be an interesting research topic. 

The natural question to be answered is under which scenarios InfoNEAT and its counterparts~\cite{related_works:automated_hyperparameter_tuning,related_works:hyperparameter_tuning_RL} should be used. 
In a nutshell, similar to these studies, training time would constitute the cost of applying InfoNEAT. 
Since we trained submodels one by one, training took 2 days on average (including training of the stacked model), comparable to 4 days and approximately 3.5 days reported in~\cite{related_works:hyperparameter_tuning_RL} and~\cite{related_works:automated_hyperparameter_tuning}, respectively.  
The stacking technique used in InfoNEAT facilitates the parallelization of the algorithm, as discussed next. 

\vspace{5pt}\noindent\textbf{Parallelization of InfoNEAT:}
Generally, NEAT and its variants can enjoy the benefits offered by parallel evolutionary algorithms to optimize memory allocation and time complexity~\cite{cheng2019accelerating}. 
Compared to NEAT, InfoNEAT enhanced by integrating a stacked model (converting a multi-class task to a number of binary classification problems) can greatly benefit from parallelization. 
In fact, InfoNEAT fits the purpose of distributed learning in asynchronous mode~\cite{bengio2013deep} since submodels can be trained individually and in parallel.  
The results presented in this paper, however, are not obtained using any specific parallelization. 
Running parallel InfoNEAT can be considered as future work.  
For SCA, our experiments, however, suggest that InfoNEAT is scalable to the number of classes, i.e., 256 classes (see Appendix~C for an example of InfoNEAT's time complexity in our experiments). 

}
\section{Conclusion}\label{sec:conclusion}
In this paper, we introduce a first of its kind, information theory-based automatic neuroevolution technique that successfully profiles the SCA traces of $256$ different classes. This is achieved through use of the CMI-heuristic, NEAT algorithm, OvA decomposition technique, and  consequent stacking ensemble learning to train an effective SCA model which is simple, compact, and yet deep enough to successfully extract the sub-key using less than a hundred traces. 
\newversion{In addition, we provide a detailed study 
on how InfoNEAT can be implemented, trained, and tested in practice. 
The effectiveness and feasibility of InfoNEAT are demonstrated by applying it against widely-used side-channel datasets.}
\ndss{For future work, particular research directions will be considered, including the application of InfoNEAT against other datasets, e.g.,~\cite{ASCAD_v2,port_dataset}. 
Furthermore, the stacking technique used in InfoNEAT facilitates the parallelization of the algorithm, which has not yet been explored by us and is left for future work. }

\section*{Acknowledgment} 
The authors would like to thank Mr. Mohammad Hashemi and Dr. James Kingsley (Worcester Polytechnic Institute) as well as the HyperGator Team (University of Florida) for their support and help with the experimental setup. 
Results in this paper were obtained in part using a high-performance computing system acquired through NSF MRI grant DMS-1337943 to WPI. 
This study has also been supported in part by NSF under award number 2138420, Air Force's Center of Excellence for Enabling Cyber Defense in Analog and Mixed Signal Domain (CYAN) under the contract number FA8650-19-1-1741, and AFOSR MURI Grant under Award FA9550-14-1-0351.  

\small
\bibliographystyle{alphaurl}
\bibliography{references}

\vspace{20pt}
\appendix 
\noindent\textbf{{Appendix A. Related Work at a Glance}}
\vspace{5pt}

Table~\ref{tab:related_works} summarizes some of the most relevant studies discussed in detail in Section~\ref{sec:related_works}. 
\ndss{Moreover, Table~\ref{tab:related_NAS} compares InfoNEAT with existing studies considering NAS for SCA. }

\begin{table*}[h]
\centering
\caption{Table showing the most recent and state-of-the-art works in the field of SCA. The table also elaborates on the problem investigated as well as the comparison to our approach, InfoNEAT. 
In this table, RF, SVM refer to random forest and support vector machine algorithms, respectively. \vspace{5pt}}
\label{tab:related_works}
\resizebox{\linewidth}{!}{%
\begin{tabular}
{|p{0.07\textheight}|
p{0.43\textheight}|
p{0.05\textheight}|
p{0.43\textheight}|}
\hline
\multicolumn{1}{|c|}{\textbf{Ref.}} & \multicolumn{1}{c|}{\textbf{Problem investigated}} & \multicolumn{1}{c|}{\textbf{Method}} & \multicolumn{1}{c|}{\textbf{Comparison to our approach (InfoNEAT)}} \\ 

\hline
 \cite{related_works:explainability_SCA} & Explains why MLP works, how the internal nodes are working, and how different they are compared to other models in terms of the internal representation & MLP & This work ends up using a smaller network and smaller training dataset similar to our approach. \\ 

\hline
 \cite{related_works:plaintext}& Uses the plaintext feature to make the ML models more powerful. Claims that CNN is most successful because of their effectiveness with raw data. & CNN & 
 Without considering the plaintext, our approach along with~\cite{related_works:explainability_SCA,related_works:ensemble_SCA,related_works:MLP_weissbart} have been quite effective and comparable to the works using CNN. \\ 

\hline
 \cite{related_works:MI_SCA}& 
 Discusses a stopping criterion based on the analysis of mutual information. & MLP & 
 We use Matrix-based R\'enyi's $\alpha$-entropy CMI rather than usually applied mutual information to define stoping criterion. 
 Network architecture and other hyperparameters are evolved automatically.  \\ 

\hline
 \cite{hettwer2020applications} & Surveys the state-of-the-art techniques used for the purpose of SCA. And explains that most of the techniques including CNN and MLP are quite comparable when it comes to SCA.
 & CNN, MLP & In line with this survey, we use our methodology to train compact MLP-like networks.\\ 


\hline
 \cite{related_works:SCA_metrics}& Discusses the metrics that should be used for evaluating the models developed for the purpose of SCA. & MLP & We use the same metrics especially the average rank to evaluate our InfoNEAT model.\\ 

\hline
 \cite{related_works:one_trace_CNN}& Shows how all ML techniques are effective in attacking the EdDSA, especially CNN which was able to break the implementation with a single measurement. & RF, SVM, CNN & They attack a different dataset compared to us, but nevertheless, show that only few traces are required for an effective SCA.\\ 

\hline
 \cite{related_works:CNN_performance}& Evaluates the performance of CNN and MLP for the purpose of SCA. Also shows that considering the minor performance gains that CNN offers, it is not worthy to invest time and resource to design such a complex network. & CNN, MLP & We therefore use InfoNEAT to design simple and compact MLP-like networks and show that they are quite effective.\\ 

\hline
 \cite{ascad_paper}& Introduces the ASCAD dataset structured in a similar way as compared to a typical ML dataset. & CNN, MLP & We use the ASCAD dataset to evaluate our model and our results are comparable to the ones presented in this paper.\\
 
\hline

 \cite{related_works:MLP_weissbart}& Evaluates the performance of MLP and shows that they are quite effective for the purpose of SCA.& MLP & We also show that MLP-like network is very effective for SCA.\\ 
 \hline

 \cite{related_works:ensemble_SCA}& Explains that output class probabilities are sensitive to small changes and thus the developed model is not generalizable. They use ensemble of different models to create a generalizable model. & MLP & We also use ensembles of models (stacking) to develop an effective model. Our stacked model involves submodel for each class, devised using the One-vs-All classification technique. And the whole training process is automated.\\ 
 
 \hline
 \cite{related_works:automated_hyperparameter_tuning}& This work tunes the hyperparameters using Bayesian optimization and random search.& CNN, MLP &  
 Usual techniques are used to output a set a hyperparameters, where each set used to train a model. 
 The mechanism to find the best model (i.e., without underfitting or overfitting) is partially automatic (i.e., searching a pre-defined set of parameters), in contrast
 to InfoNEAT. Our approach starts from a minimal size
 network and gradually evolves the network size and parameters automatically based on cross-entropy loss and CMI-values. \\
 
 \hline
  \cite{related_works:hyperparameter_tuning_RL}& Talks about the problems related to traditional deep learning models- they are complex and there are too many hyperparameters to tune. & CNN & This work uses reinforcement learning (RL) techniques to tune the hyperparameters of CNN.\\ 

 \hline
 
\end{tabular}
}
\end{table*}

\begin{table*}[h]
\centering
\caption{\ndss{
Different aspects of NAS considered in SCA-related studies and InfoNEAT. (\textcolor{mygreen}{\checkmark}, \textcolor{mygreen}{\checkmark*}, and 
\textcolor{red}{\ding{55}} are used to denote a task that has been completely, partially (e.g., with a range defined for the hyperparameter values), and by no means performed, respectively. } \vspace{5pt}}
\label{tab:related_NAS}
\resizebox{\linewidth}{!}{%
\begin{tabular}
{|c|c|c|c|c|}
\hline
\multicolumn{1}{|c|}{\textbf{Ref.}} & \multicolumn{1}{c|}{\textbf{Max. no. of layers}} & \multicolumn{1}{c|}{\textbf{Type of operation}} & \multicolumn{1}{c|}{\textbf{Hyperparameters}} & \multicolumn{1}{c|}{\textbf{Connection config.}} \\ 
\hline
\cite{related_works:MI_SCA} & \textcolor{red}{\ding{55}} & \textcolor{red}{\ding{55}} & \textcolor{mygreen}{\checkmark*} & \textcolor{red}{\ding{55}}\\

\hline
\cite{perin2020influence}& \textcolor{mygreen}{\checkmark*} & \textcolor{mygreen}{\checkmark} & \textcolor{mygreen}{\checkmark} & \textcolor{red}{\ding{55}}\\ 

\hline
\cite{kerkhof2021no}& \textcolor{mygreen}{\checkmark*} & \textcolor{mygreen}{\checkmark} & \textcolor{mygreen}{\checkmark} & \textcolor{red}{\ding{55}}\\

\hline
\cite{zaid2020methodology}& \textcolor{mygreen}{\checkmark*} & \textcolor{mygreen}{\checkmark} & \textcolor{mygreen}{\checkmark} & \textcolor{red}{\ding{55}}\\

\hline
 \cite{related_works:MLP_weissbart}& \textcolor{red}{\ding{55}} & \textcolor{mygreen}{\checkmark} & \textcolor{mygreen}{\checkmark} & \textcolor{red}{\ding{55}}\\ 
 \hline

 \cite{related_works:ensemble_SCA}& \textcolor{red}{\ding{55}} & \textcolor{mygreen}{\checkmark} & \textcolor{mygreen}{\checkmark} & \textcolor{red}{\ding{55}}\\ 
 
 \hline
 \cite{related_works:automated_hyperparameter_tuning}& \textcolor{mygreen}{\checkmark*} & \textcolor{mygreen}{\checkmark} & \textcolor{mygreen}{\checkmark} & \textcolor{red}{\ding{55}} \\
 
 \hline
  \cite{related_works:hyperparameter_tuning_RL}&  \textcolor{red}{\ding{55}} & \textcolor{mygreen}{\checkmark} & \textcolor{mygreen}{\checkmark} & \textcolor{red}{\ding{55}} \\ 

 \hline
 
\textbf{InfoNEAT} &  \textcolor{mygreen}{\checkmark} & \textcolor{mygreen}{\checkmark} & \textcolor{mygreen}{\checkmark} & \textcolor{mygreen}{\checkmark} \\ 

 \hline
\end{tabular}
}
\end{table*}

\vspace{10pt}

\noindent\textbf{{Appendix B. Network and Configuration Parameters}}

\begin{table}[t]
\centering
\caption{\majorrevision{InfoNEAT hyper/parameters, mainly set to the default values defined in NEAT. }}
\label{tab:NEAT_parameters_neat}
\resizebox{0.8\columnwidth}{!}{%
\begin{tabular}{|c|c|c|} 
\hline
\textbf{Parameters} & \textbf{Description} & \textbf{Values} \\ 
\hline
\begin{tabular}[c]{@{}c@{}}fitness\\threshold~($L_{TH}$)\end{tabular} & \begin{tabular}[c]{@{}c@{}}When the fitness value meets or exceeds this value, \\the algorithm halts.\end{tabular} & 0.0 \\ 
\hline
\multicolumn{3}{|c|}{\textbf{Network parameters}} \\ 
\hline
\begin{tabular}[c]{@{}c@{}}initial \\num\_hidden~($n_h$)\end{tabular} & \begin{tabular}[c]{@{}c@{}}Refers to the initial number of hidden nodes in the\\network.\end{tabular} & 10 \\ 
\hline
\begin{tabular}[c]{@{}c@{}}Maximum number \\of generations~($T$)\end{tabular} & \begin{tabular}[c]{@{}c@{}}Refers to the maximum number of generations the~\\algorithm runs until the fitness criteria is not met. Usually,\\the algorithm converges before it reaches $T$ generations.\end{tabular} & 30 \\ 
\hline
\begin{tabular}[c]{@{}c@{}}Connection\\add probability~($P_c$)\end{tabular} & \begin{tabular}[c]{@{}c@{}}The probability that mutation will add a connection\\between existing nodes.\end{tabular} & 0.8 \\ 
\hline
\begin{tabular}[c]{@{}c@{}}Node add\\probability~($P_n$)\end{tabular} & \begin{tabular}[c]{@{}c@{}}The probability that mutation will add a new node,\\essentially replacing an existing connection with a node.\end{tabular} & 1.0 \\ 
\hline
\begin{tabular}[c]{@{}c@{}}Activation~\\function\end{tabular} & \begin{tabular}[c]{@{}c@{}}Part of every node and helps calculate the output of the\\node when given an input or a set of inputs.\end{tabular} & \begin{tabular}[c]{@{}c@{}}Leaky\\ReLU\end{tabular} \\ 
\hline
\begin{tabular}[c]{@{}c@{}}Fitness~\\function\end{tabular} & \begin{tabular}[c]{@{}c@{}}Guides the evolution process within the NEAT framework.\end{tabular} & \begin{tabular}[c]{@{}c@{}}Log\\loss\end{tabular} \\

\hline
\end{tabular}
}
\end{table}

Here we expand on why and how some of the parameters in NNs are set in the InfoNEAT algorithm, which is tuned to launch key-recovery attacks against side-channel traces. 
Besides parameters set by the user and summarized in Table~\ref{tab:NEAT_parameters}, some parameters given in Table~\ref{tab:NEAT_parameters_neat} remain in their default setting as defined in the NEAT algorithm. 

\vspace{1ex}

\noindent\textbf{The fitness function and the fitness threshold: }
One of the major functions that guides the evolution process within the NEAT framework is the objective function or commonly known as the \textit{fitness function}. The fitness function which is specified by the user is used to evaluate the quality of the solution or the genome. 
In the case of InfoNEAT, we have selected \textit{logistic loss} (hereafter called log loss) as our fitness function as shown in Section~\ref{sec:neat}. 
This is due to the fact that calculating log loss can give the same quantity as calculating the cross-entropy. 
Specifically, log loss is one of the most commonly used loss functions which measures the performance of a classification model whose output is a probability value ranging from $0$ to $1$. 
The lower the log loss value, the better the model is. 
With the fitness function selected, we must also specify the fitness threshold, the value when if reached halts the evolutionary algorithm. This threshold value is set as $0$ as specified in Table~\ref{tab:NEAT_parameters}. 

\noindent\textbf{Network parameters and hyperparameters:} \ndss{The following hyperparameters can be considered when running NEAT. 

\emph{Initial number of hidden nodes} ($n_h$) indicates how many hidden neurons are involved in the NNs in the first generation. 
NEAT starts out minimally to ensure that the solution in the lowest-dimensional weight space is searched first. 
This further enhances the time-complexity since smaller structures optimize faster than larger structures~\cite{stanley2002evolving}. 
For this, the NNs in the first generation have one hidden layer with the number of neurons $n_h$. 
Over the generations, the number of neurons in each layer increases until the stopping criteria are fulfilled. 
These criteria include the \emph{maximum number of generations} defined to stop NEAT if other criteria are not fulfilled. 
In our experiments, InfoNEAT always stops before reaching to this number of generations, thanks to the CMI-based criterion. 

To determine with what probability the mutation operation adds a connection between existing nodes, \emph{connection add probability} ($P_c$) can be set. 
This probability is not changed over the generations. 
The larger this probability is set, the more diverse range of NNs can be evolved. 
In doing so, the larger number of species and greater topological diversity lead to obtaining a larger population, although the per generation training time would increase. 
Nevertheless, we observe that $P_c=0.8$ has been suitable in all our experiments. 
Similarly, \emph{node add probability} ($P_n$) indicates the probability of inserting a new node in an existing connection. 
The high $P_n$ encourages NEAT to explore new architecture per generation; hence, it is set to its maximum, i.e., $P_n=1$. 
This probability remains the same over generations.}

One of the most important network parameters is the \emph{activation function}. 
In this paper, we have selected the \textit{Leaky ReLU} or Leaky Rectified Linear Unit as our activation function for all the nodes (except the output nodes) rather than the widely used ReLU activation function (see Table~\ref{tab:NEAT_parameters}). This was done to prevent the well-known ``dying ReLU'' problem -- the scenario when a large number of the nodes output zero because of their inputs being negative, which causes the network to saturate early and train very slowly~\cite{activation_functions}. Furthermore, Leaky ReLU has been shown to have better performance compared to ReLU and other activation functions such as tanh, sigmoid, etc.~\cite{activation_functions}. 

\vspace{10pt}


\begin{figure}[t]
    \centering
    \subfloat[\centering]{{\includegraphics[width=0.47\textwidth]{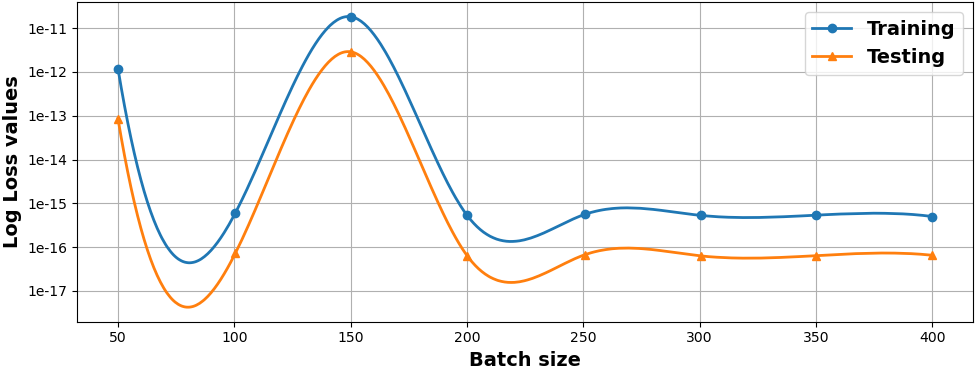} }\label{fig:results_HE_initialization}}
    \hfill
    \subfloat[\centering]{{\includegraphics[width=0.47\textwidth]{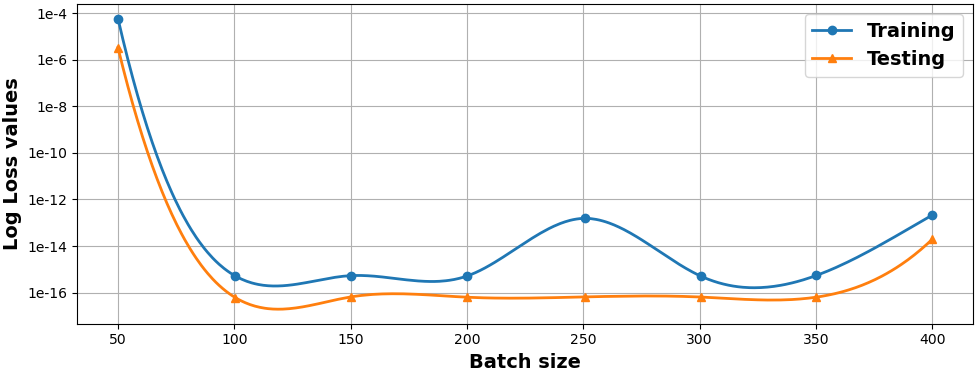} }\label{fig:results_xavier_initialization}}
    \caption{Log loss values for different batch sizes. The weights and biases are initialized based on (a) He, and (b) Xavier initialization techniques. The results are shown for a randomly chosen class.}
    \label{fig:results_weights_intialization}
\end{figure}

\vspace{1ex}

\noindent\textbf{Initialization of weights and batch size for training:}
To prevent convergence issues, including the \textit{saturation problem}~\cite{kumar2017weight}, we look at two initialization techniques, namely He and Xavier. 
Figures~\ref{fig:results_HE_initialization} and~\ref{fig:results_xavier_initialization} show the results of using the aforementioned initialization techniques in terms of log loss values for models trained and tested with different batch sizes. 
Note that we select the batch data from the dataset randomly, and we make sure that the data is balanced and unbiased. 
Moreover, for the sake of readability, we present these results for one class (sub-key), although similar observations are made for other classes.  
In this case, we use the same batch size for training and testing. 
Based on our results, the He initialization can lead to underfitting (see the curve in Figure~\ref{fig:results_HE_initialization} for batch size being greater than 250). 
Moreover, a sharp minimum is observed when batch size is equal to 100 that can result in poor generalization~\cite{keskar2016large}. 
Although it can be thought that the Xavier initialization is not a suitable method for networks with ReLU activation function, as our MLPs are not considerably deep, Xavier initialization can be as powerful as He initialization~\cite{kumar2017weight}; hence, we choose Xavier initialization. 

\vspace{1ex}

\begin{figure}[t]
\centering
\includegraphics[width=0.8\columnwidth]{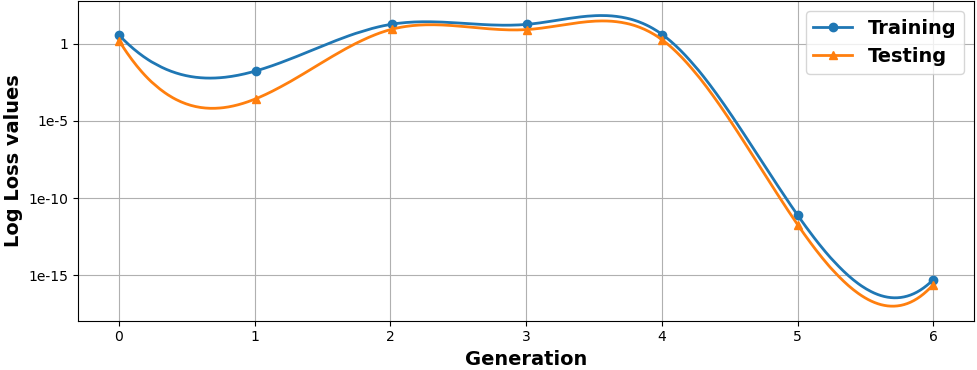}
\caption{Learning curve showing the log loss values for different generations. The training and the testing curve values are obtained using the best trained model against the corresponding training and testing dataset in each generation respectively.
The result is obtained for a randomly chosen class and the batch size is set to 150.}
\label{fig:results_learning_curve}
\end{figure}

\noindent\textbf{Batch size selection: }To select the \textbf{batch size}, one has to make a trade-off between the size of the batches and the speed of training and generalization error. 
On the one hand, reducing the size of the batches can have a noise-like effect that is useful for regularizing the data. 
On the other hand, choosing a larger batch size has been considered as a solution to speed up the training by parallelizing computations. 
In contrast to this, as reported in~\cite{morse2016simple}, EAs can run significantly faster by using smaller batch sizes. 
Hence, to keep the generalization gap as small as possible, we focus on the batch sizes $100$ and $150$. 
To speed up the training, we measure the average time spent to evolve the genomes before the algorithm halts, which is $653\,s$ and $582\,s$ for batch size equals to $100$ and $150$, respectively. 
Therefore, we use the batch size of $150$. 
To observe the impact of this, the diagnostic learning curves are drawn, where the Xavier initialization technique is applied to train the model. 
The fittest genome or network in each generation is then tested against the test data, and the corresponding log loss values are plotted in Figure~\ref{fig:results_learning_curve}. The figure shows that the models trained with InfoNEAT are neither overfitting nor underfitting. This is a good sign that the trained model will be generalizable (see  Section~\ref{sec:results_sca_metrics} for more detail). 
Additionally, the selection of Xavier initialization is further justified based on this result.

\vspace{1ex}

\noindent\textbf{{Appendix C. Examples Illustrating the Time and Memory-efficiency of InfoNEAT }}

\begin{figure*}[t]
\centering
\includegraphics[width=1\textwidth]{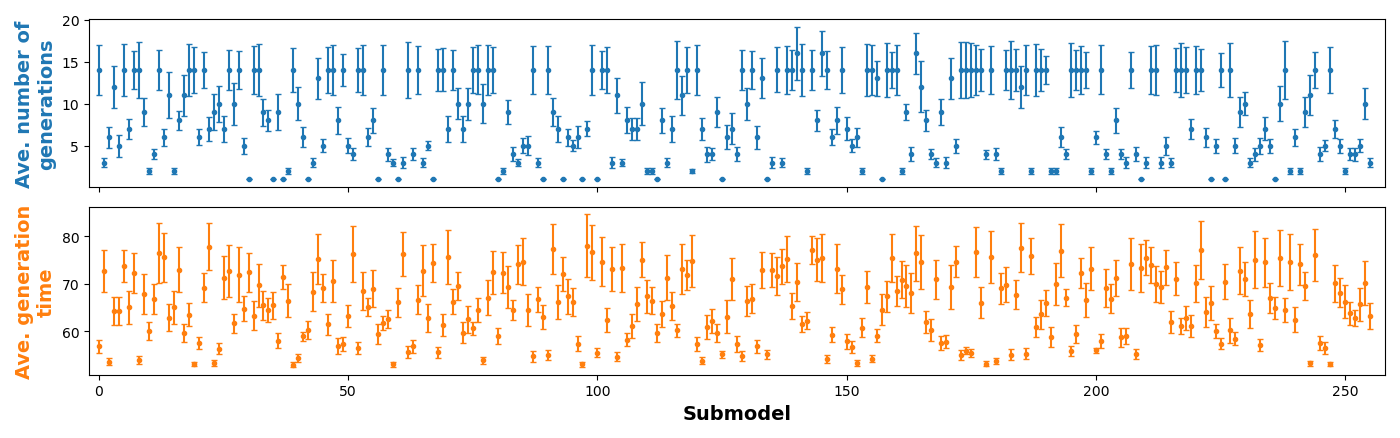} 
\caption{An example of the average number of generations (top) and generation time [s] (bottom) for all the $256$~sub-models trained on ASCAD\SPSB{fixed}{sync}. 
These values are obtained from across all ten folds to be trained. 
The error bars represent the Standard Error Mean (SEM) with $95\%$ confidence interval. }
\label{fig:results_average_gen_time}
\end{figure*}

\vspace{5pt}\noindent\textbf{Time-complexity: }
\newversion{To further elaborate on the time complexity of InfoNEAT, Figure~\ref{fig:results_average_gen_time} illustrates the average time spent per generation and the average number of generations for each sub-key in ASCAD\SPSB{fixed}{sync} (a similar trend has been observed for ASCAD\SPSB{fixed}{desync50} and ASCAD\SPSB{fixed}{desync100}). 
As can be understood from the graphs, although a set of NNs including the species and genomes is trained per generation, the training remains feasible thanks to the CMI-based stopping criterion. 
Interestingly, per submodel, the average training time (536.95s) is much less than what has been reported in~\cite{ascad_paper} (5475s) cf.~\cite{related_works:automated_hyperparameter_tuning}, and its order is comparable to the studies focused on hyperparameter tuning (405s)~\cite{related_works:automated_hyperparameter_tuning}. 
Nevertheless, InfoNEAT offers additional benefits as discussed in Section~\ref{sec:related_work_hyper}, and as a trade-off, the training time is slightly increased. 
}

\begin{figure*}[t]
\centering
\includegraphics[width=1\textwidth]{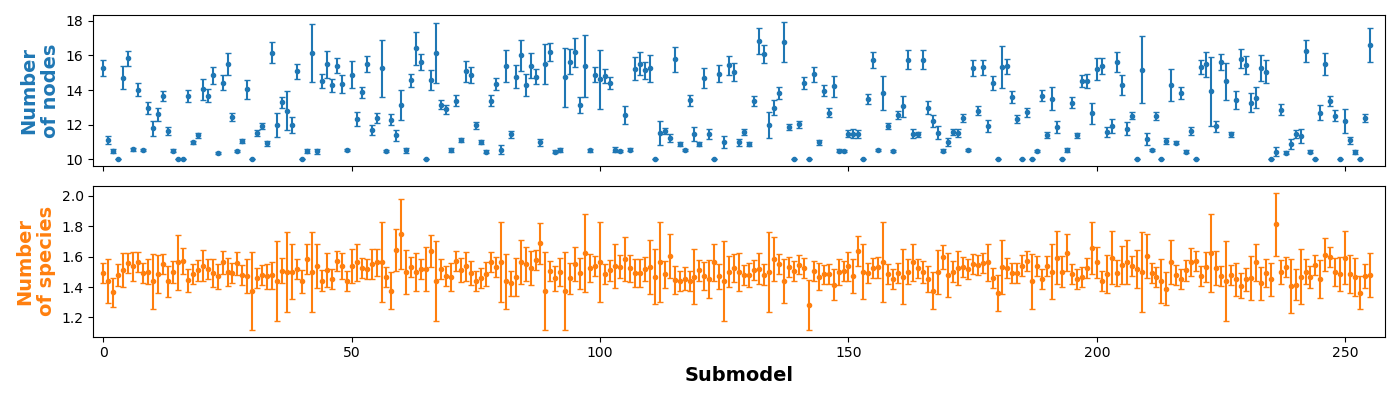} 
\caption{The number of nodes (top) and species (bottom) for all the $256$~sub-models.  These values are obtained for all $N$ genomes across different generations throughout the training process. The error bars represent the SEM with $95\%$ confidence interval.}
\label{fig:results_number_nodes_species}
\end{figure*}

\vspace{5pt}\noindent\textbf{Memory-efficiency: }
\newversion{Additional information on the number of nodes and number of species for all classes (corresponding to $256$ sub-keys and sub-models) is depicted in Figure~\ref{fig:results_number_nodes_species}. 
Note that although the genomes are relatively small, they perform well and recover the sub-key efficiently (see Figures~\ref{fig:results_neat_all_dataset} and \ref{fig:results_neat_comparison}). 
}

\end{document}